\documentclass[fleqn,usenatbib]{mnras}

\usepackage{graphicx}	% Including figure files
\usepackage{amsmath}	% Advanced maths commands
\usepackage{amssymb}	% Extra maths symbols
\usepackage{multicol}        % Multi-column entries in tables
\usepackage{pdflscape}	% Landscape pages
\usepackage{lipsum}

\usepackage{newtxtext,newtxmath}
\usepackage[T1]{fontenc}
\DeclareRobustCommand{\VAN}[3]{#2}
\let\VANthebibliography\thebibliography
\def\thebibliography{\DeclareRobustCommand{\VAN}[3]{##3}\VANthebibliography}
\usepackage{xspace}	
\usepackage{color}
\usepackage{lineno}
\usepackage{fleqn,natbib}
\usepackage{flushend}
\usepackage{cuted}
\usepackage{rotating}
\usepackage{xcolor}
\usepackage{dcolumn}
\usepackage{multicol}
\usepackage{adjustbox}
\usepackage{hyperref}
\newcommand{\thisgrb}{GRB~211211A\xspace}
\newcommand{\thisgrbo}{GRB~220910A\xspace}
\newcommand{\thisgrboa}{GRB~180720B\xspace}
\newcommand{\thisgrbob}{GRB~181222B\xspace}
\newcommand{\fermi}{{\em Fermi}\xspace}

\newcommand{\sw}[1]{\texttt{#1}}
\graphicspath{{./}{figures/}}

\usepackage{color}
\usepackage{amsmath}
\usepackage{multirow,longtable,supertabular}
\usepackage{float}
\usepackage{tablefootnote}
\usepackage{threeparttable}

\title[X-ray and gamma-ray timing of GRBs]{X-ray and gamma-ray timing of GRB~180720B, GRB~181222B, GRB~211211A and GRB~220910A observed with {\it Fermi} and ASIM}

\author[Caballero-Garc\'{i}a et~al.]{M.~D. Caballero-Garc\'{i}a$^{1}$,\thanks{Email: mcaballero@iaa.es}
E. G{\"o}{\v{g}}{\"u}{\c{s}}$^{2}$,
J. Navarro-Gonz\'alez$^{3}$,
K. E. Atapin$^{4}$,
E. Sonbas$^{5,6}$,
 \newauthor
M. Uzuner$^{2}$, 
A. J. Castro-Tirado$^{1,7}$,
S. B. Pandey$^{8}$, 
Rahul Gupta$^{9}$,
A. K. Ror$^{8}$,
Y.-D. Hu$^{10}$, 
\newauthor
S.-Y. Wu$^{1}$,  
R. S\'anchez-Ramirez$^{1}$, 
S. Guziy$^{1}$, 
F. Christiansen$^{11}$,  
P. H. Connell$^{3}$,
T. Neubert$^{11}$, 
  \newauthor
N. {\O}stgaard$^{12}$, 
J. E. Adsuara$^{3}$,
F. J. Gordillo-V\'azquez$^{1}$, 
E. Fern{\'a}ndez-Garcia$^{1}$, \newauthor 
I. P\'erez-Garcia$^{1}$, and
V. Reglero$^{3}$
\\
$^{1}$ Instituto de Astrof\'isica de Andaluc\'ia (IAA-CSIC), Glorieta de la Astronom\'ia s/n, E-18008, Granada, Spain\\
$^{2}$ Sabanc\i~University, Faculty of Engineering and Natural Sciences, \.Istanbul 34956 Turkey \\
$^{3}$ Image Processing Laboratory, University of Valencia, Paterna, Valencia, Spain \\
$^{4}$ Sternberg Astronomical Institute, Moscow State University, Universitetsky pr., 13, Moscow, 119991, Russia \\
$^{5}$ Department of Physics, Adiyaman University, 02040 Adiyaman, Turkey \\
$^{6}$ Department of Physics, The George Washington University, Washington, DC 20052, USA \\
$^{7}$ Unidad Asociada al CSIC, Departamento de Ingenier\'ia de Sistemas y Autom\'atica, Escuela de Ingenier\'ias, Universidad de M\'alaga, M\'alaga, Spain \\
$^{8}$ Aryabhatta Research Institute of Observational Sciences (ARIES), Manora Peak, Nainital-263002, India \\
$^{9}$ Astrophysics Science Division, NASA Goddard Space Flight Center, Mail Code 661, Greenbelt, MD 20771, USA \\
$^{10}$ INAF-Osservatorio Astronomico di Brera, Via E. Bianchi 46, I-23807 Merate (LC), Italy \\
$^{11}$ National Space Institute, Technical University of Denmark, Kgs. Lyngby, Denmark \\
$^{12}$ Birkeland Centre for Space Science, Department of Physics and Technology, University of Bergen, Norway \\
}

\pubyear{2024}

\begin{document}
\label{firstpage}
\pagerange{\pageref{firstpage}--\pageref{lastpage}}
\maketitle

\begin{abstract}
We present a timing study of the gamma and X-ray observations and analysis of a sample of bright gamma-ray bursts (GRBs; i.e. GRB~180720B, GRB~181222B, GRB~211211A and GRB~220910A), including the very bright and long \thisgrb (a.k.a. kilonova candidate). They have been detected and observed by the Atmosphere-Space Interactions Monitor (ASIM) installed on the {\it International Space Station} ({\it ISS}) and the Gamma-ray Burst Monitor (GBM) on-board the \fermi mission. The early (${\rm T}-{\rm T}_{0}{\approx}$\,s) and high-energy (0.3-20\,MeV) ASIM High Energy Detector (HED) and (150\,keV-30\,MeV) {\it Fermi} (BGO) light curves show well-defined peaks with a low quasi-periodic oscillation (QPO) frequency between $2.5-3.5$\,Hz that could be identified with the spin of the neutron star in the binary mergers originating these GRBs. These QPOs consist on the first detection of low-frequency QPOs (${\le}10$\,Hz) detected in magnetars so far. We also detect a strong QPO at $21.8-22$\,Hz in GRB~181222B together with its (less significant) harmonics. The low-frequency QPO would correspond to the signal of the orbiting neutron star (NS) previous to the final coalescence giving rise to the gravitational-wave (GW) signal.
\end{abstract}


%%%%%%%%%%%%%%%%%%%%%%%%%%%%%%%%%%%%%%%%%%%%%%%%%%

%%%%%%%%%%%%%%%%% BODY OF PAPER %%%%%%%%%%%%%%%%%%

% The MNRAS class isn't designed to include a table of contents, but for this document one is useful.
% I therefore have to do some kludging to make it work without masses of blank space.
%\begingroup
%\let\clearpage\relax
%\tableofcontents
%\endgroup
%\newpage
%\vspace{-3.0}

\clearpage
\newpage

\section{Introduction}

\noindent Gamma-ray bursts (GRBs) are the most energetic and catastrophic explosions in the Universe after the Big Bang. They are classified into two categories depending on their duration, i.e. short and long GRBs (sGRB and lGRB) with durations of $<$ and $>$ 2s, respectively. 
The lGRBs have been associated with the collapse of massive stars. Otherwise the sGRBs are commonly believed to be powered by the accretion of a massive remnant disc onto the compact black hole (BH) or neutron star (NS) remnant following the compact binary merger. The consequent thermal, novae-like transient (kilonova) gives rise to the radioactive decay of heavy, neutron-rich elements synthesized in the expanding merger ejecta \citep{2019MNRAS.489.2104T}. 

The GRB-associated gravitational wave event GRB 170817A/GW170817 \citep{2017ApJ...848L..12A} and a few kilonovae-associated GRBs \citep{2013Natur.500..547T,2015NatCo...6.7323Y} have added new clues to the origin of GRBs. Nevertheless the study of the electromagnetic (EM) counterpart of GW events is necessary for their understanding. Of particular importance is
the study of their {\it precursors}, i.e. the EM emission from the astrophysical merger before its collapse. Currently the association between sGRBs and kilonovae has extended to lGRBs as well. Some studies of a collection of sGRBs and lGRBs \citep{2019MNRAS.485.5294P,2022ApJ...931L..23L} concluded that some lGRBs have signatures of kilonovae-like progenitor contrary to any previous expectation from this kind of sources.

GRB 211211A is a kilonova-associated gamma-ray burst whose light curve consists of a precursor, a hard spiky emission and a soft long extended emission (with a duration of ${\approx}0.2,10,40$\,s, respectively) which has attracted great attention. \citet{2022arXiv220410864R} reported the discovery of a kilonova associated with this nearby (350\,Mpc) minute-duration GRB 211211A confirmed later by \citet{2022Natur.612..228T}. Kilonova association could prove its merger origin, while the detection of the precursor could be used to infer at least one highly magnetized neutron star (NS) being involved in the merger. \citet{2022ApJ...934L..12G} report that the special behavior of GRB 211211A is mainly due to the strong magnetic field of its progenitor star. 
%\par

It has been proposed that in the late in-spiral phase of a NS-NS or NS-BH system in which one of the components is a magnetar (NS), the tidal force on the magnetar due to its companion would increase dramatically as the components of the binary approach. The tidal-induced deformation may surpass the maximum that the crust of the magnetar could sustain just seconds or sub-seconds before the coalescence. A catastrophic global crust destruction could then occur, and the magnetic energy stored in the interior of the magnetar would be released thus being observed as a super-flare with energy hundreds of times larger than the giant flares of magnetars, thus a sGRB \citep{2022ApJ...939L..25Z}. 

A few studies of the fast X-ray variability in GRBs have been performed in order to reveal the so elusive Quasi-Periodic Oscillations (QPOs) in GRBs \citep{2021ApJ...911...20T,2024JCAP...07..070L}. Since QPOs are associated to the innermost regions around compact objects any significant detection of them has the potential to give important insights into their physical origin.

In this paper we give further clues on the nature of a sample of bright GRBs (GRB~180720B, GRB~181222B, GRB~211211A and GRB~220910A) through the analysis of their fast-time X-ray and gamma-ray variability observed with ASIM \citep{2019arXiv190612178N} and {\it Fermi}\footnote{\url{https://fermi.gsfc.nasa.gov}}. We first start putting these GRBs in context in Sec. 1.1. In Section 2 we report on the timing observations and analysis made with ASIM and {\it Fermi} where we detect significant and simultaneous QPOs. Finally in Section 3 we discuss the results obtained and the implications of the discovery of such QPOs.

\begin{figure}
\centering
\includegraphics[angle=0,scale=0.50]{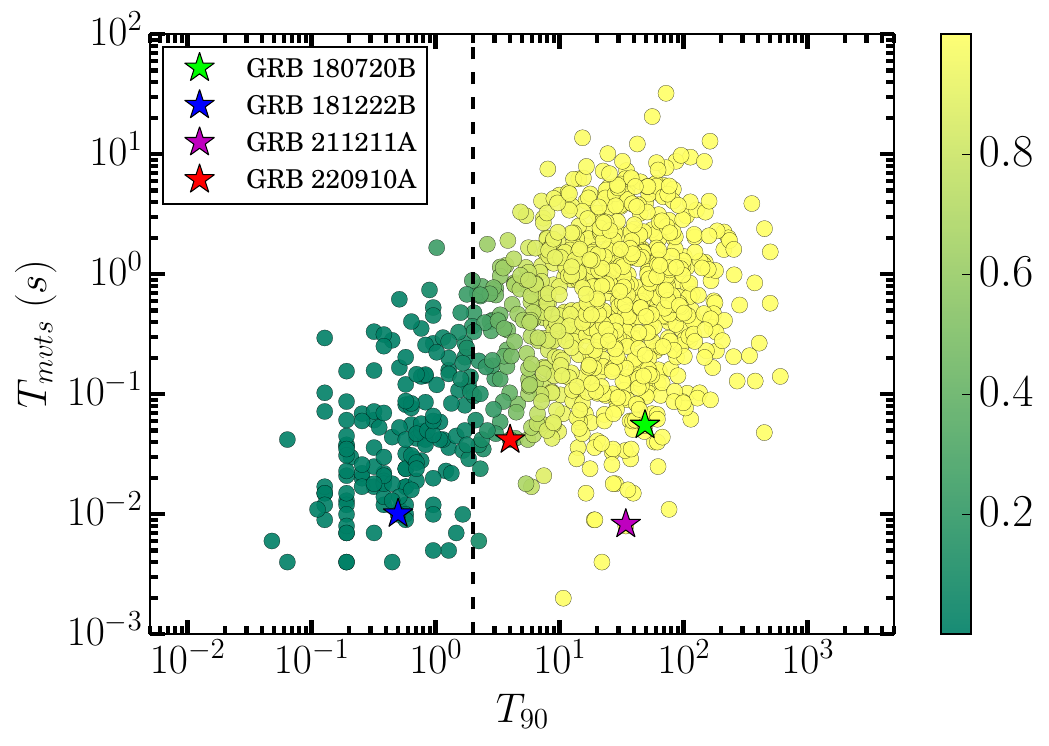}
\includegraphics[angle=0,scale=0.65]{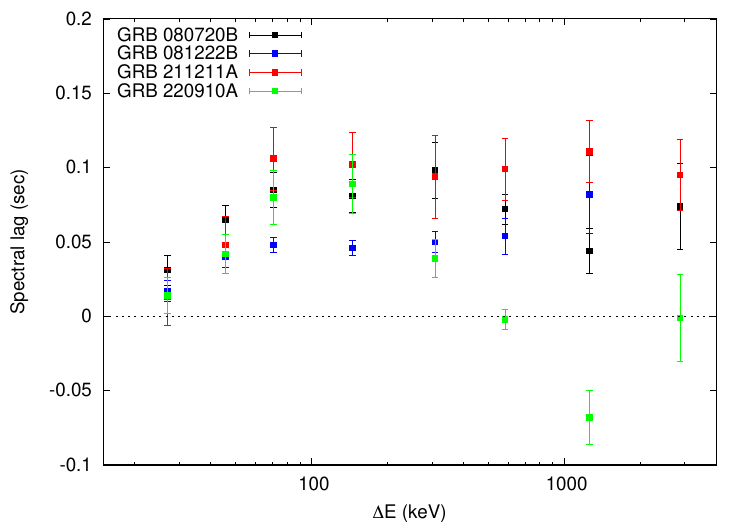}
\caption{(Upper) Minimum variability timescale for each burst using the Bayesian blocks binding method (together with data from \citealt{2015ApJ...811...93G}). The probabilities of each GRB originating from the merger (by fitting a gaussian mixture model to the minimum variability time scale) versus T90 distribution are also shown. (Lower) Spectral lag as a function of energy for various observes frame energy bands extracted using GBM light curves in QPO signal.} 
\label{mvts}
  %\vspace{-0.3cm}
\end{figure}

%\vspace{-0.3cm}

\subsection{Minimum variability time scale and spectral lag}

Minimum variability time scale (MTS/MVTS) is the smallest temporal feature in the light curve that is consistent with a fluctuation above the Poisson noise level and potentially provides a quantitative means of probing the regional size of the emission location involved. Methods for extracting such variabilities using a technique based on wavelets are well described \citep{2012MNRAS.425L..32M, 2013MNRAS.432..857M, 2013MNRAS.436.2907M, 2014ApJ...787...90G, 2015ApJ...811...93G}. The MTS has been shown to follow several correlations involving temporal and spectral features \citep{2012MNRAS.425L..32M, 2015ApJ...805...86S, 2023A&A...671A.112C}. Here we extracted the time variability following the work of \citet{2013MNRAS.432..857M} using GBM light curves in a time range that covers the frequencies in and around the QPO signals. The extracted MTS in the QPO range is found to be systematically smaller than that found in the continuum sections of the PSD for all GRBs (Fig. \ref{mvts}). This suggests a smaller source emission size for the QPO process compared to the process that produces the continuum).

Spectral lags are defined as the arrival time differences between high- and low-energy photons and are seen in significant fraction of long duration GRBs. Hard-to-soft evolution of the spectrum produces positive spectral lags, while soft-to-hard evolution leads to negative lags. We extracted spectral lags for various observer-frame energy bands in QPO signal using the CCF method \citep{2002ApJ...579..386N, 2006Natur.444.1044G, 2008ApJ...677L..81H, 2010ApJ...711.1073U, 2012MNRAS.419..614U}. The spectral lags as a function of energy are shown in Fig. \ref{mvts}.

\begin{figure}
\centering
\includegraphics[angle=270,scale=0.29]{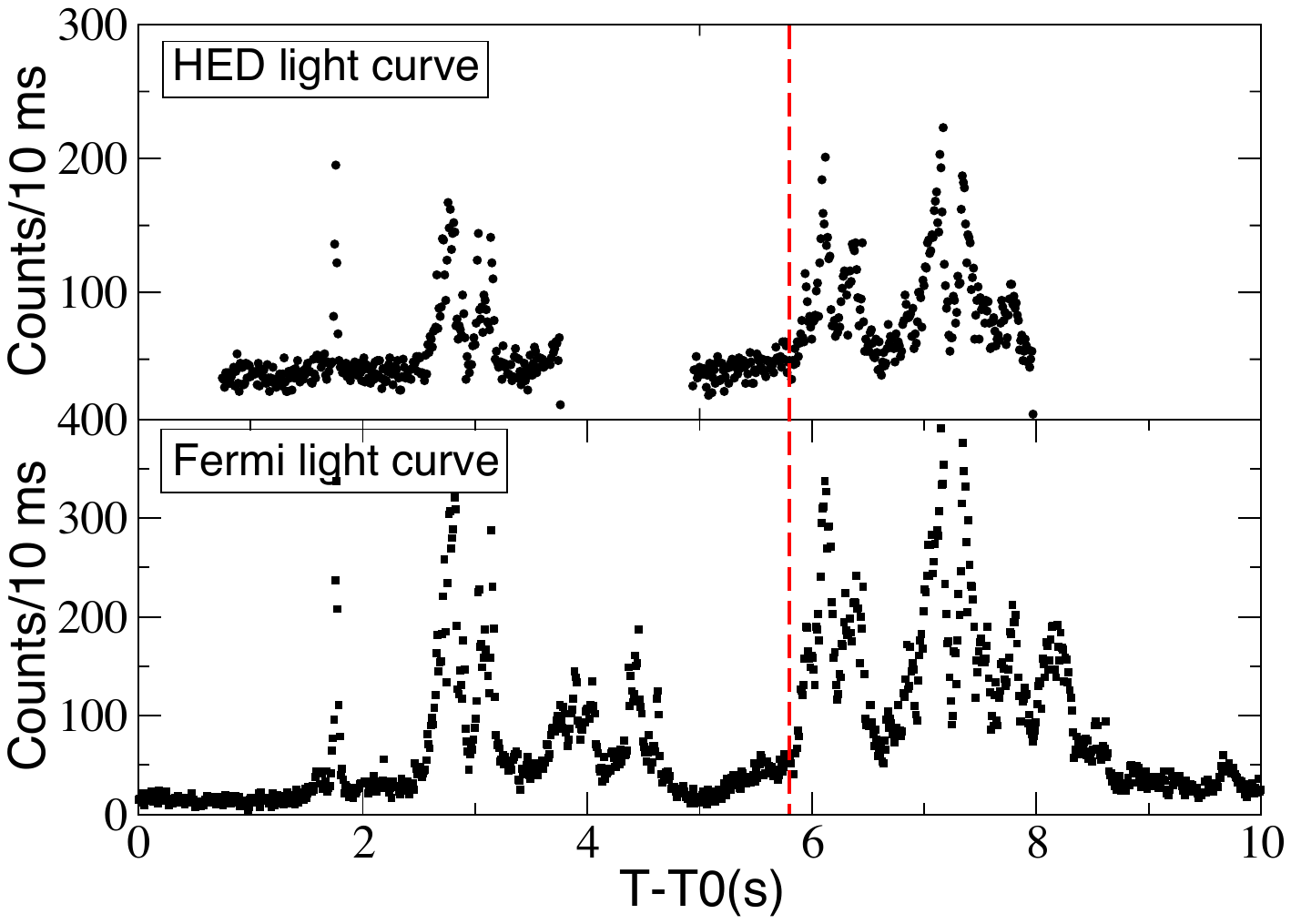}
\includegraphics[angle=270,scale=0.29]{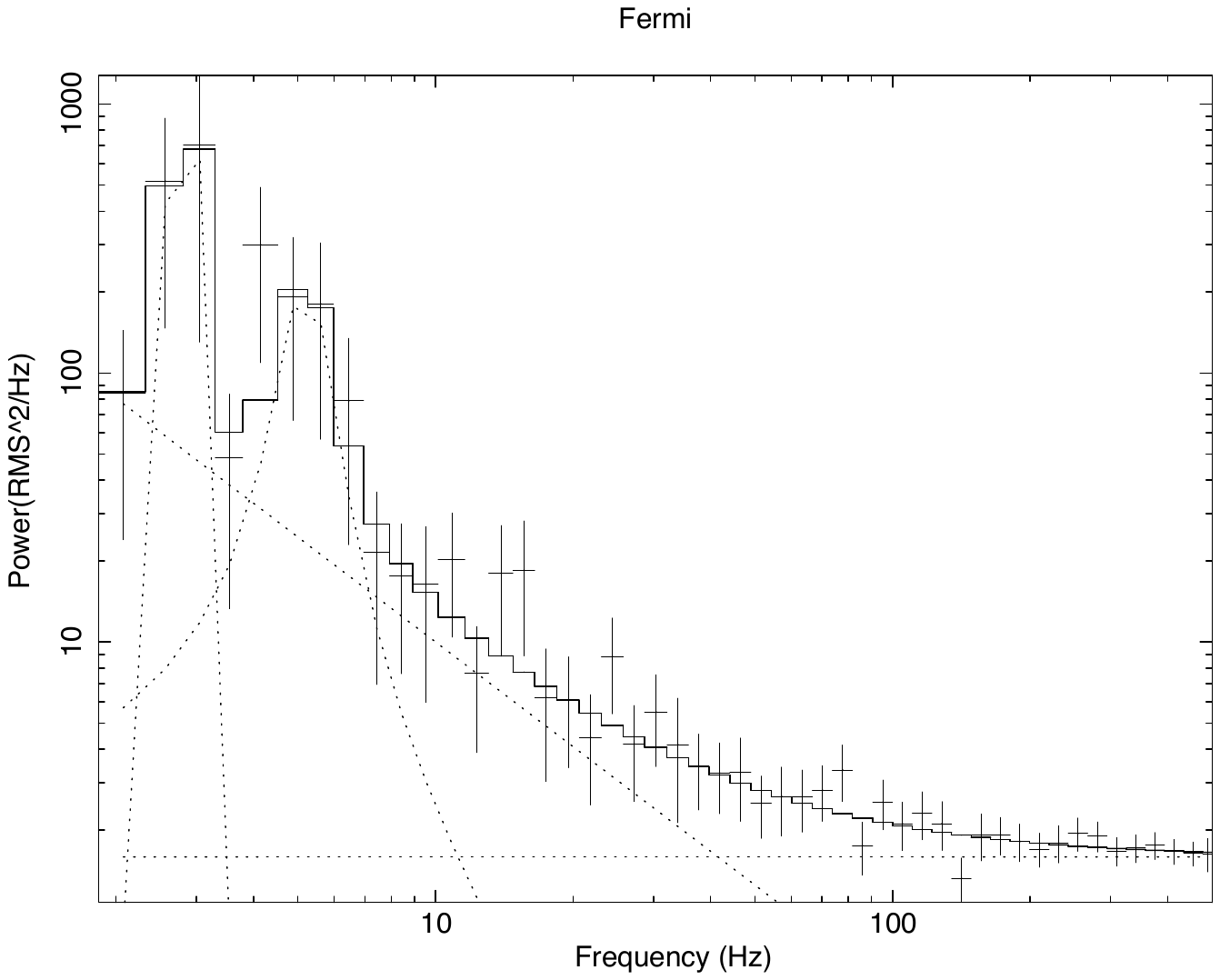}
\includegraphics[angle=270,scale=0.29]{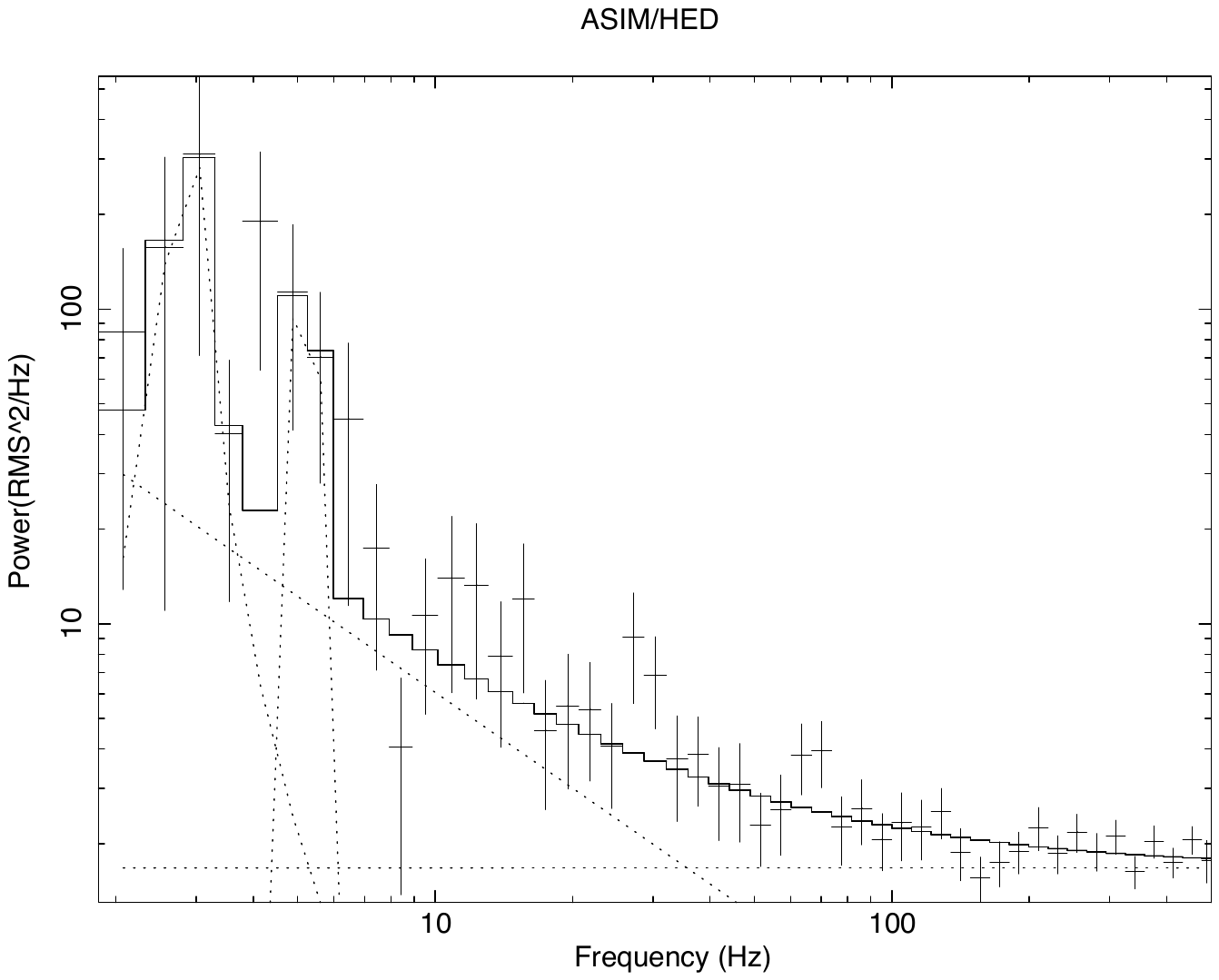}
\caption{(Top) Barycentred {\it Fermi} (BGO; 150\,keV-30\,MeV) and ASIM (HED; 0.3-20\,MeV) light curves of {\thisgrb} (${\rm T}_{0}=$13:09:59\,UT) showing the time-interval where the QPOs were the PDS was built. (Bottom) The PDS of the {\it Fermi}/b1 (8-800\,keV; upper) and ASIM/HED (0.3-20\,MeV; lower) light curves of {\thisgrb} built at times $>5.8$\,s. }
%The vertical-dashed(red) line indicates the times when the QPOs appeared.}
\label{timing1}
  %\vspace{-0.3cm}
\end{figure}

\vspace{-0.3cm}

\begin{figure}
\centering
\includegraphics[angle=270,scale=0.30]{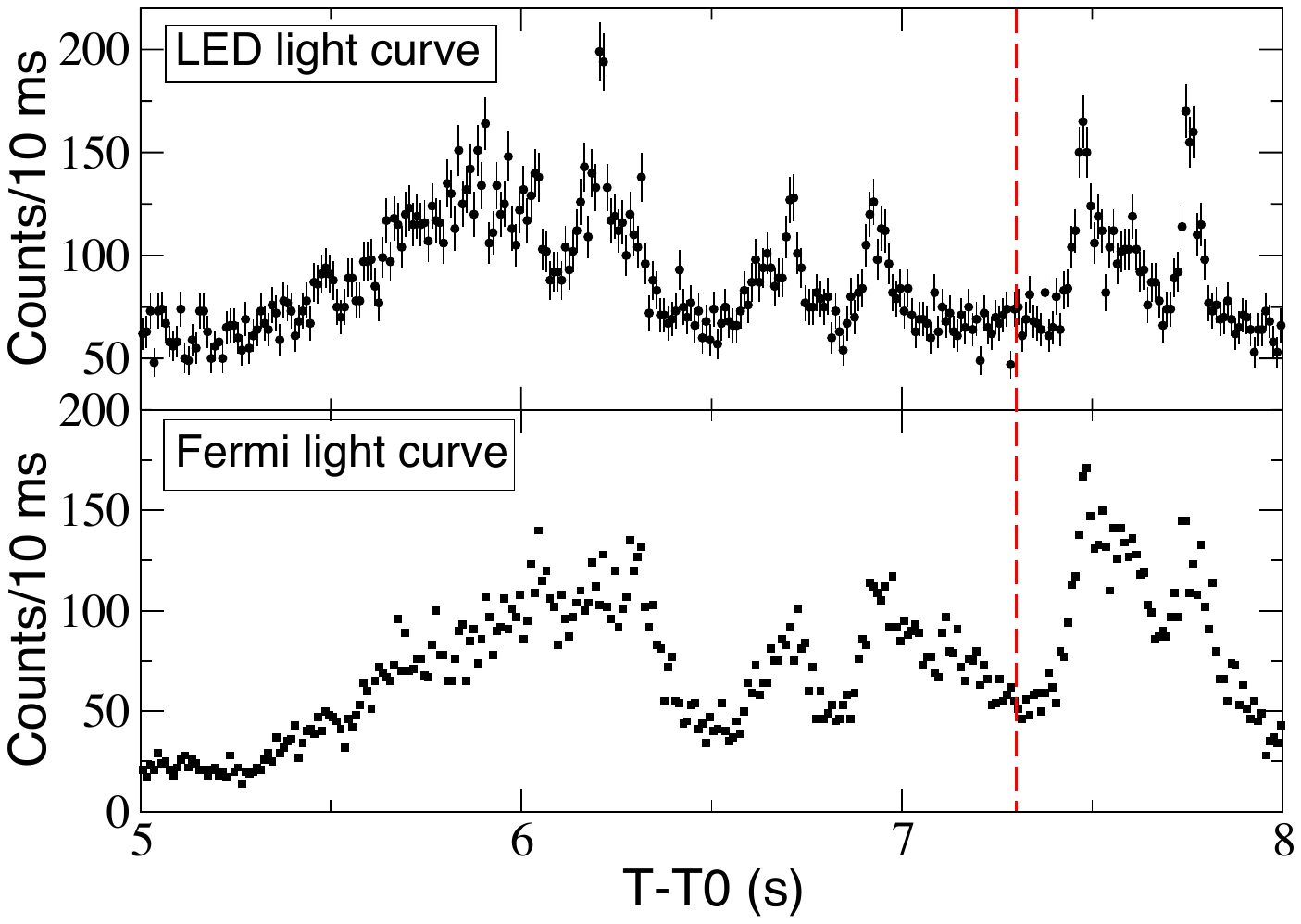}
\includegraphics[angle=270,scale=0.29]{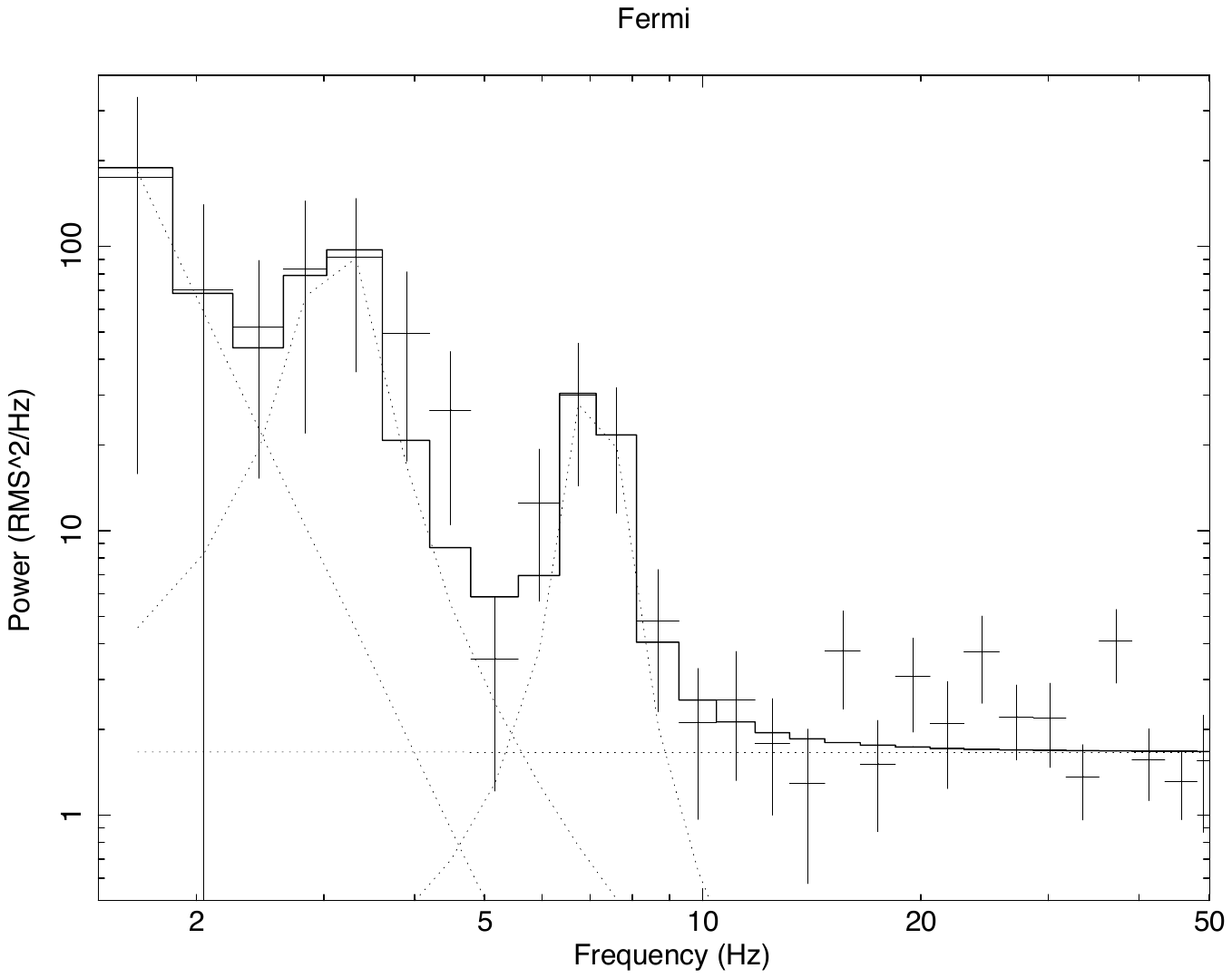}
\includegraphics[angle=270,scale=0.29]{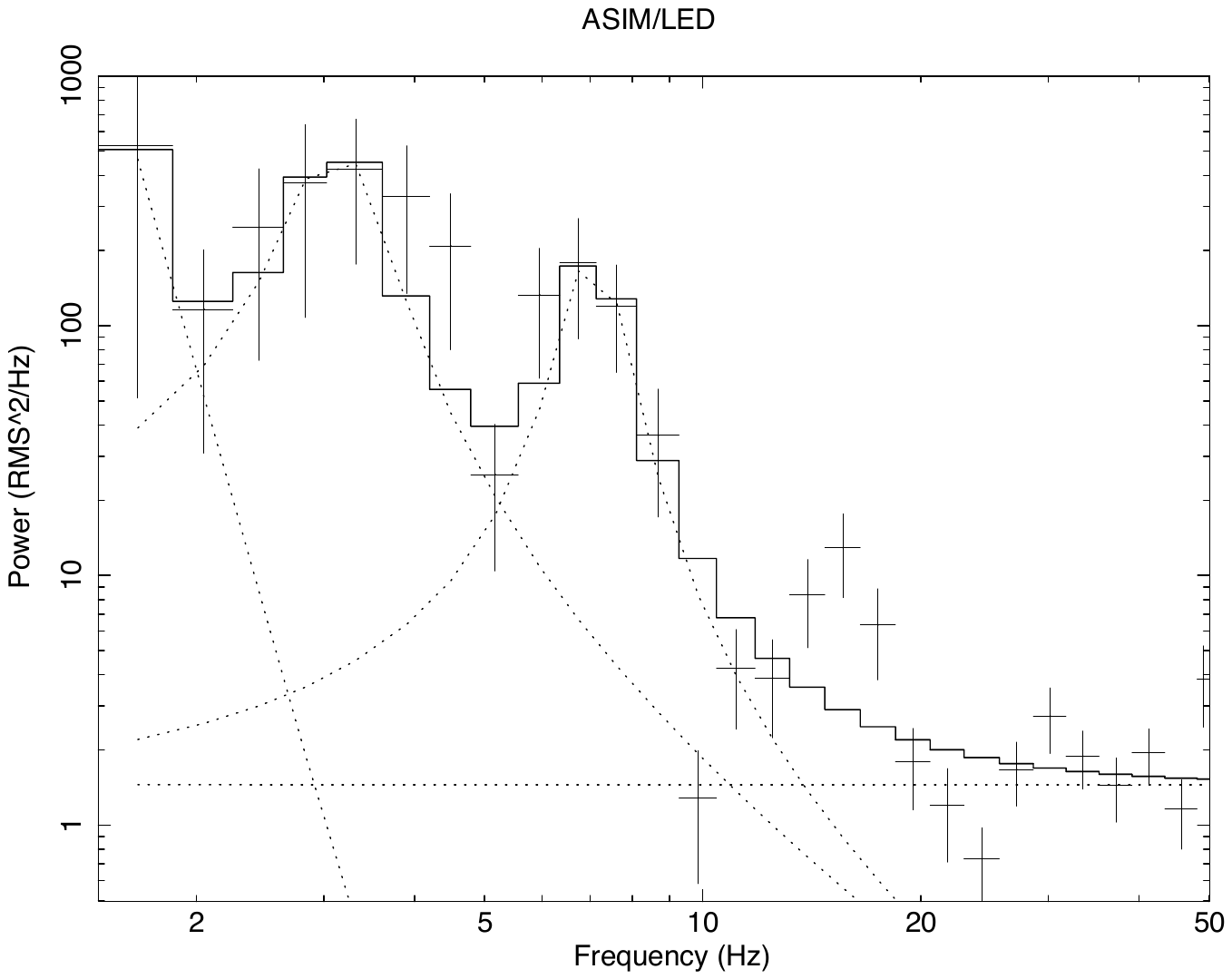}
\caption{(Top) Barycentred {\it Fermi} (NaI; 8-900 keV) and ASIM Low Energy Detector (LED; 50-400\,keV) light curves of {\thisgrbo} (${\rm T}_{0}=$05:48:21.55\,UT) showing the time-interval where the PDS was built. (Bottom) The PDS of the {\it Fermi}/nb (150\,keV-30\,MeV; upper) and ASIM/LED (50-400\,keV; lower) light curves of {\thisgrbo} built at times $>7.3$\,s.}
\label{timing2}
  %\vspace{-0.3cm}
\end{figure}

\vspace{-0.3cm}

\begin{figure}
\centering
\includegraphics[angle=270,scale=0.30]{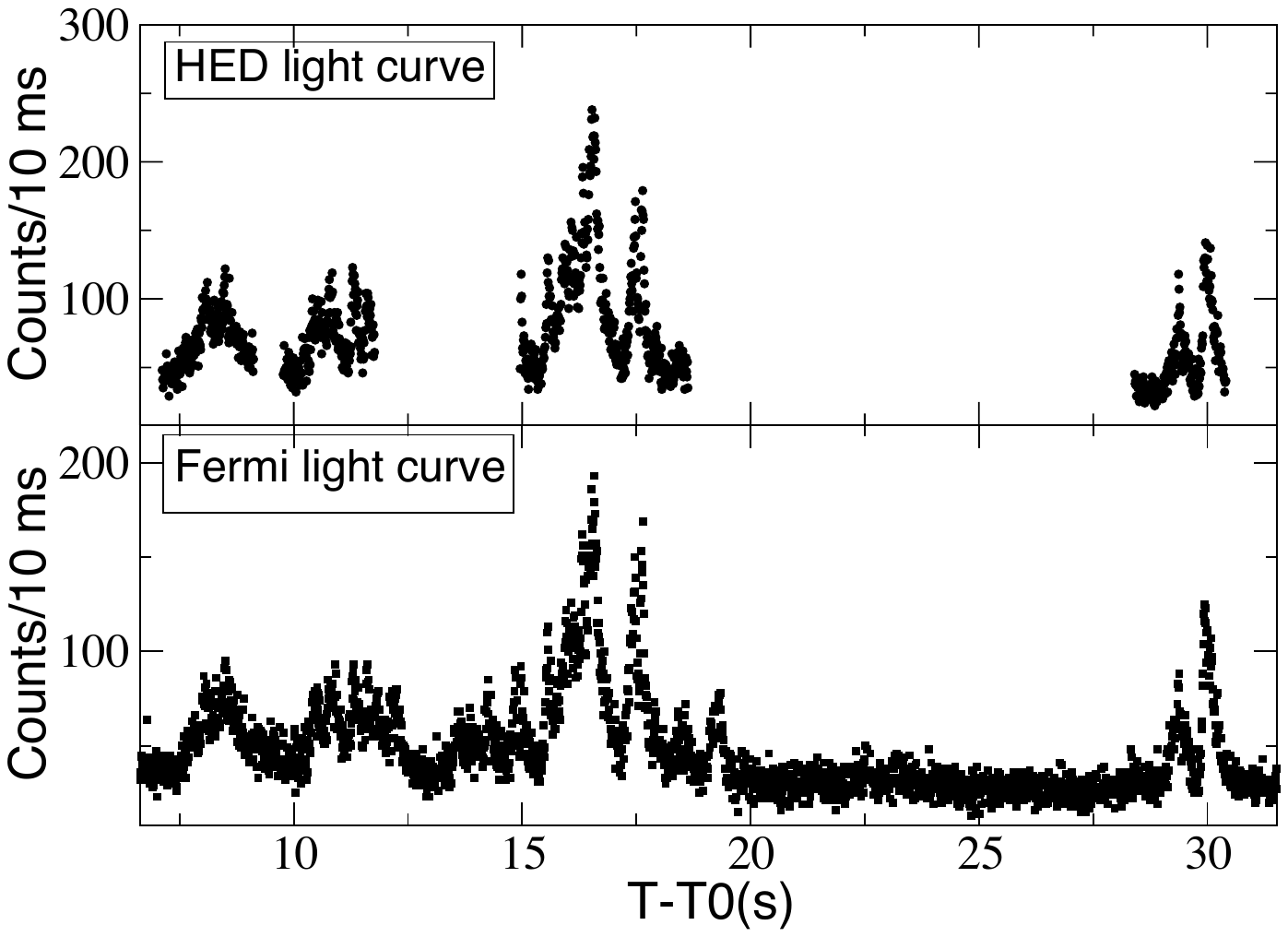}
\includegraphics[angle=270,scale=0.29]{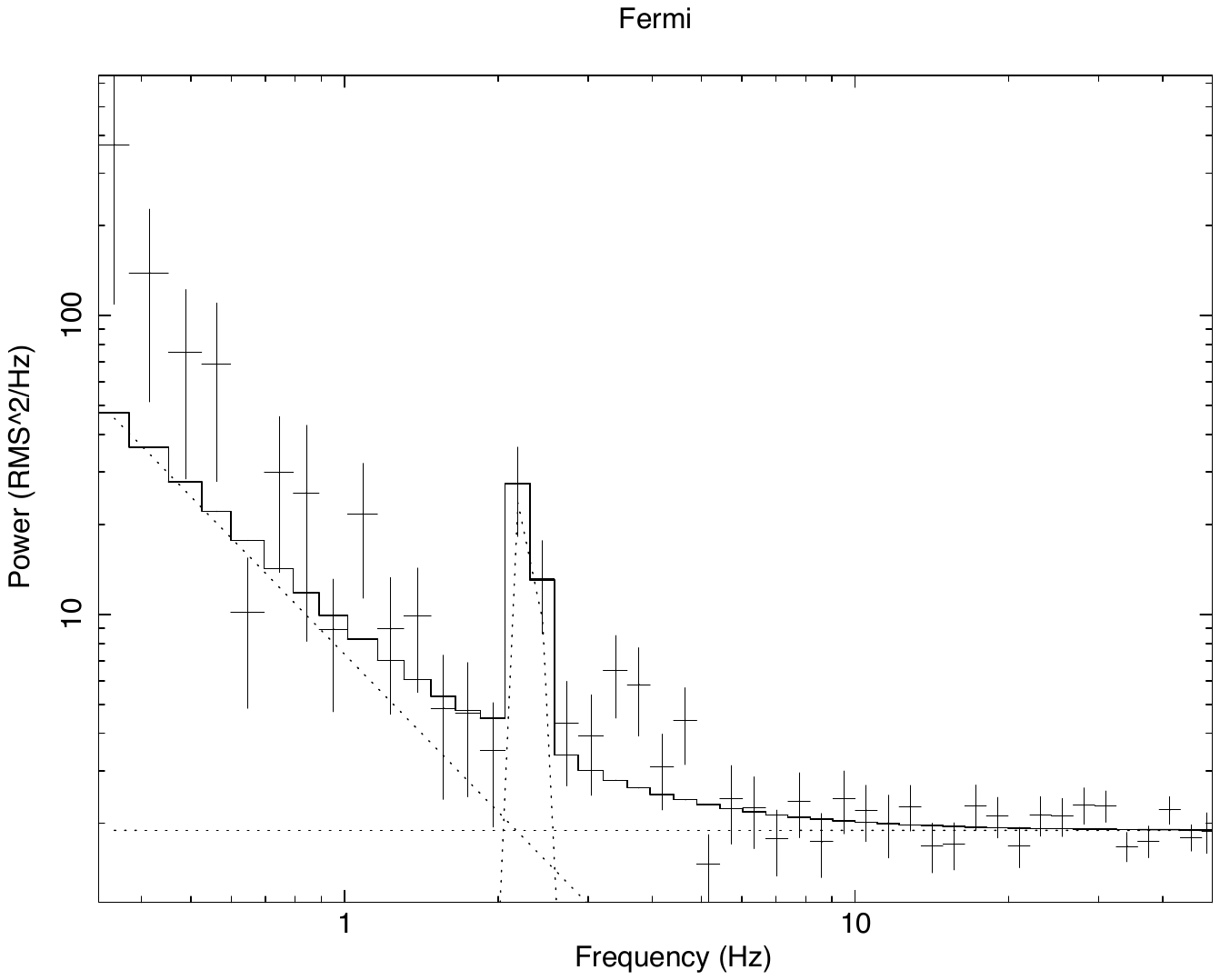}
\includegraphics[angle=270,scale=0.29]{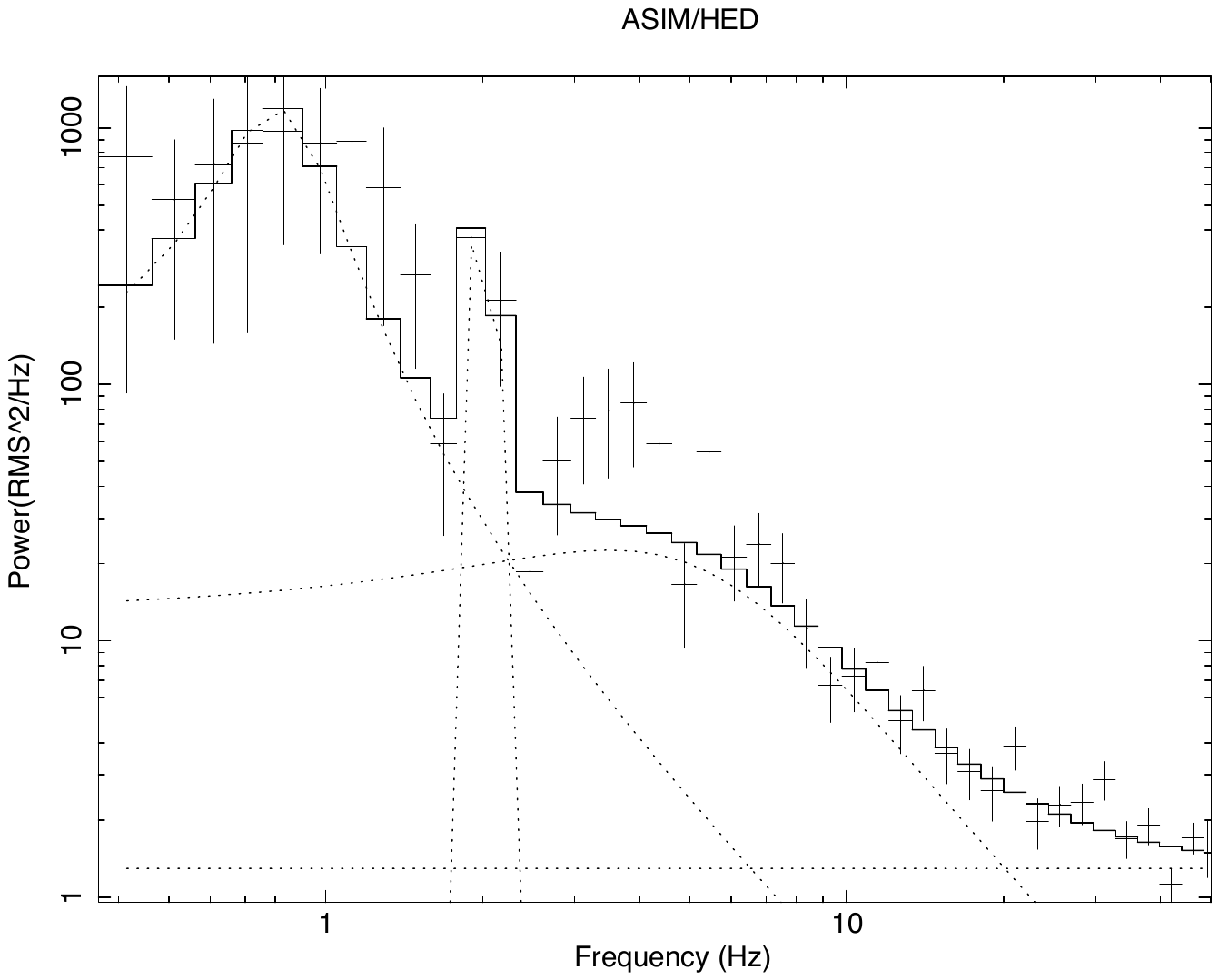}
\caption{(Top) Barycentred {\it Fermi} (BGO; 150\,keV-30\,MeV) and ASIM (HED; 0.3-20\,MeV) light curves of {\thisgrboa} (${\rm T}_{0}=$14:21:39.65\,UT) showing the time-interval where the PDS was built. (Bottom) The PDS of the {\it Fermi}/nb (8-800\,keV; left) and ASIM/HED (0.3-20\,MeV; right) light curves of {\thisgrboa} built at times ${\ge}13$\,s and ${\ge}-9-13$\,s for ASIM and {\it Fermi}, respectively.}
\label{timing3}
  %\vspace{-0.3cm}
\end{figure}

  \vspace{-0.3cm}

\begin{figure}
\centering
\includegraphics[angle=270,scale=0.30]{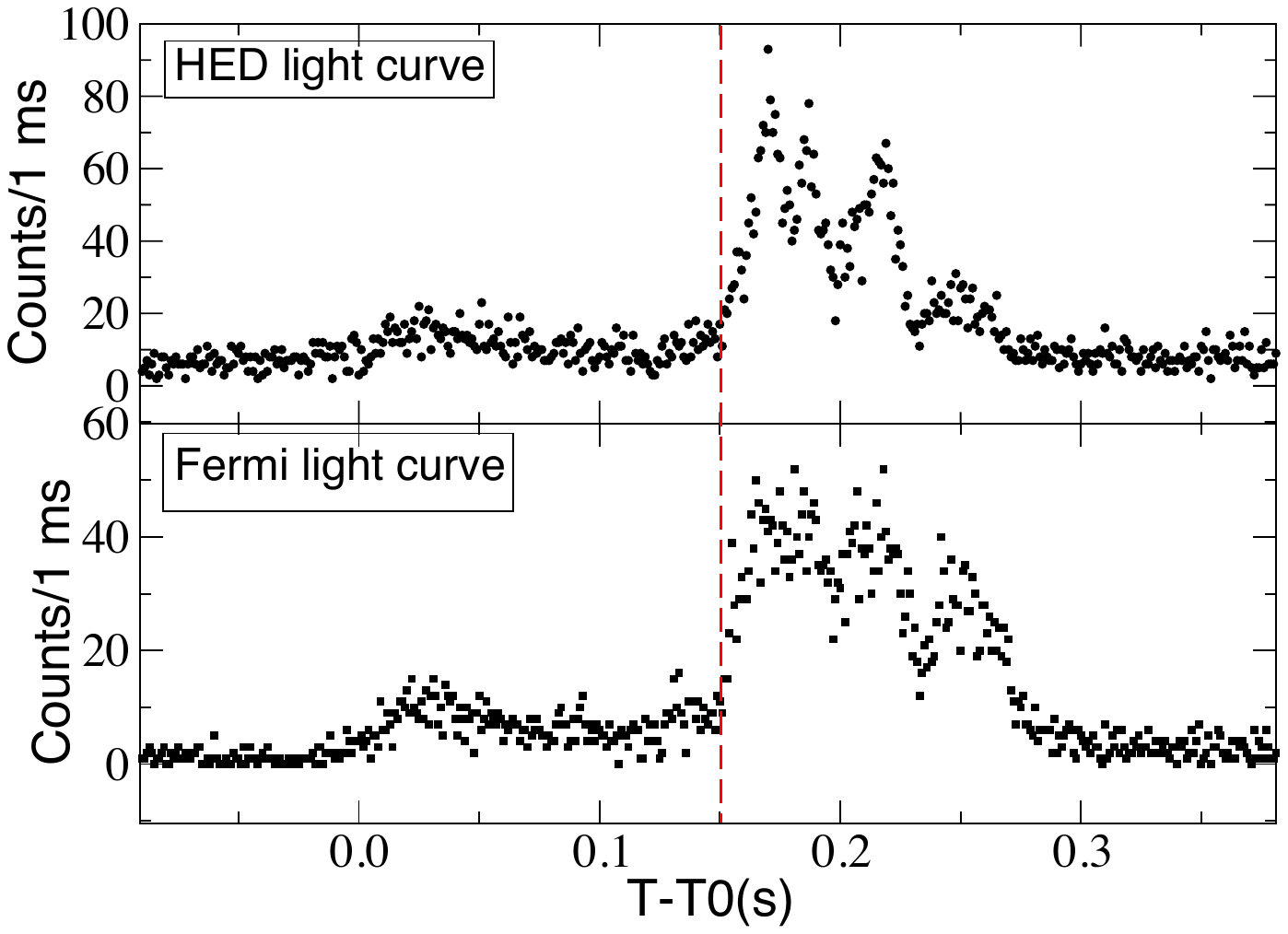}
\includegraphics[angle=270,scale=0.29]{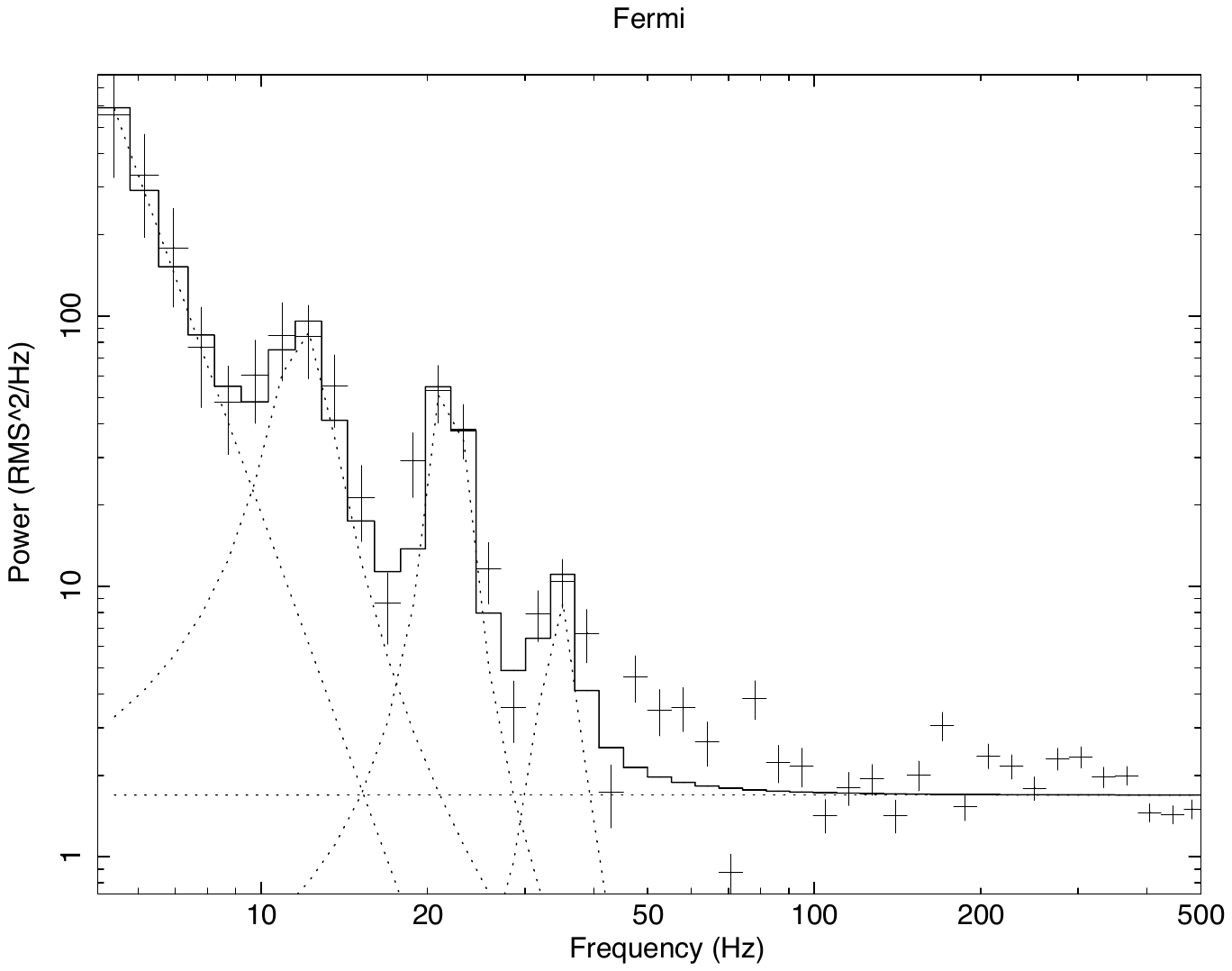}
\includegraphics[angle=270,scale=0.29]{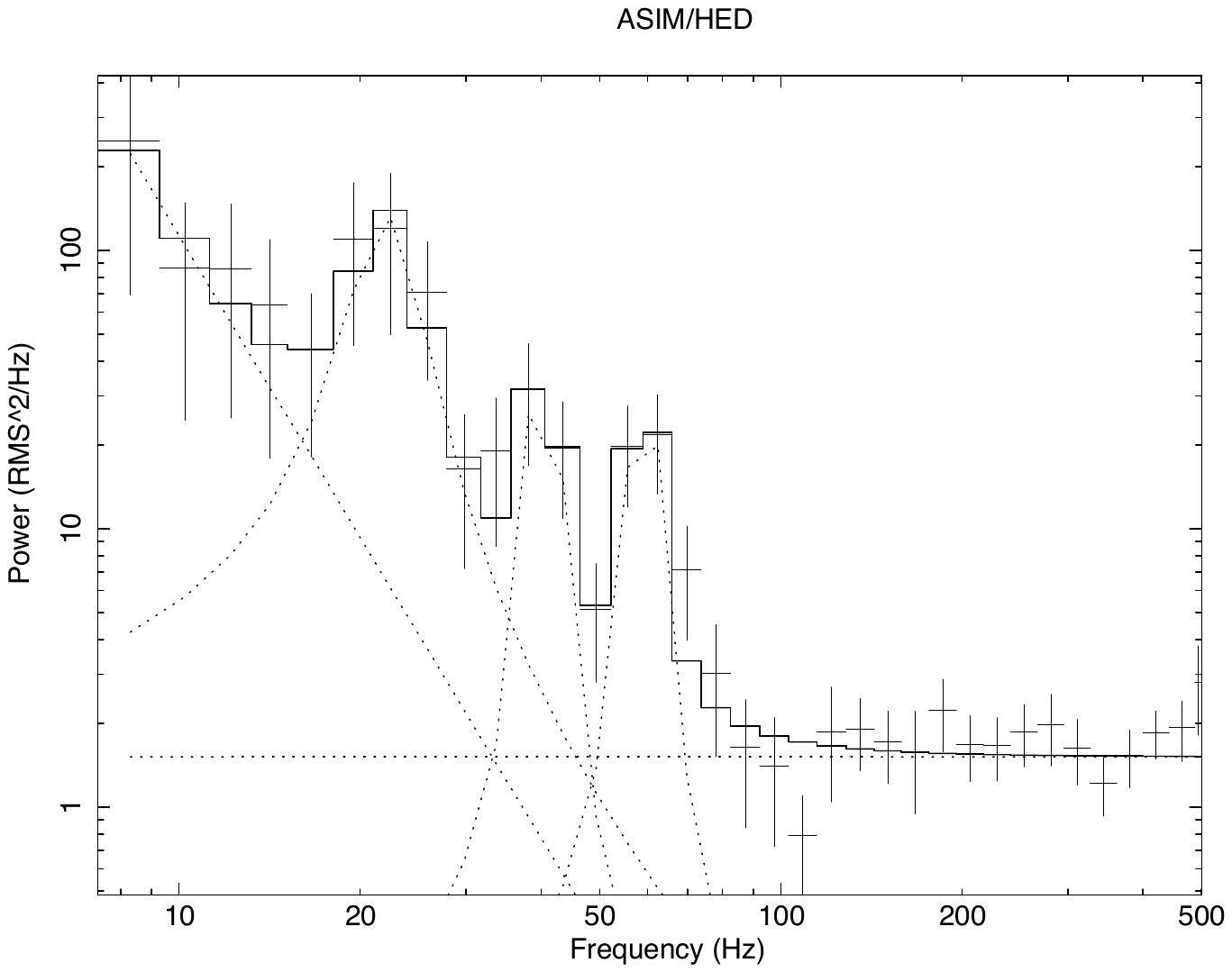}
\caption{(Top) Barycentred {\it Fermi} (BGO; 150\,keV-30\,MeV) and ASIM (HED; 0.3-20\,MeV) light curves of {\thisgrbob} (${\rm T}_{0}=$20:11:37.438\,UT) showing the time-interval where the PDS was built. (Bottom) The PDS of the {\it Fermi}/nb (8-800\,keV; left) and ASIM/HED (0.3-20\,MeV; right) light curves of {\thisgrbob} built at times $>0.15$\,s.}
\label{timing4}
  %\vspace{-0.3cm}
\end{figure}

%\vspace{-0.3cm}

%\subsection{{\it Fermi} timing analysis}

%{\bf (TODO)}

\section{Analysis and results}

The (1\,ms and 10\,ms-binned) light curves from these GRBs 
%(Fig. \ref{timing1}, \ref{timing2}, \ref{timing3}, \ref{timing4}) 
are highly variable with quasi-periodic behaviour from approximately the mid-time to the end of the ASIM and {\it Fermi} observations. We checked for this kind of variability building the Power Density Spectra (PDS) for different time intervals and found that only for certain times on-wards and during the period of activity of these GRBs the PDSs were showing significant noise in the form of QPOs. We checked the {\it Fermi} (b1) light curves as well and found the same kind of behaviour, with QPOs appearing only at the same times \footnote{Note that the ASIM and {\it Fermi} times were barycentred and referred to the {\it Fermi} BAT trigger time (i.e. ${\rm T}_{0}$ time).} (see Fig. \ref{timing1},\ref{timing2},\ref{timing3} and \ref{timing4} showing the PDSs built at these referred time-periods).

\subsection{{\it Fermi} and ASIM timing analysis of {\thisgrb} }

The PDSs of both {\it Fermi} and ASIM datasets showed the same kind of noise in the form of two low-frequency QPOs. To obtain the exact value of the peak frequency of these QPO features, the PDS was fit
with four continuum components, i.e. two power-laws, and two Lorentzians (one for each QPO).
The Lorentzian components were used to fit the QPO features (at $2.8,5.2$\,Hz for the first and second QPOs, respectively). One of the power-laws was used to fit the low frequency red noise (as done in e.g. \citealt{1990A&A...230..103B}). All these components are shown as dotted lines in Fig.\ref{timing1}-\ref{timing4} for all the GRBs.

The best PDS fit obtained had
reduced ${\chi}^2$ of $0.5,0.8$ for $40,40$ degrees of freedom for the {\it Fermi} and ASIM observations, respectively. The PDS were normalized in the Leahy Normalization \citep{1983ApJ...266..160L}. The frequency of the QPOs was the same for {\it Fermi} and ASIM observations (during the same time period). The significance of these peaks is high, i.e. significance of $5,4\,{\sigma}$ and $4,3\,{\sigma}$ (single trial and considering trials, respectively) for both ASIM/HED and {\it Fermi}. The quality factor (i.e. QPO frequency/FWHM) was of
${\rm Q}=10-500,5-500$ for the first and second peaks, respectively. The characteristics of the QPOs observed in the
PDS of the observations are listed in Table \ref{ttiming1}.

\subsection{{\it Fermi} and ASIM timing analysis of {\thisgrbo} }

As done in the case of \thisgrb to obtain the exact value of the peak frequency of these QPO features, the PDS was fit
with four continuum components, i.e. two power-laws, and two Lorentzians (one for each QPO). The Lorentzian components were used to fit the QPO features (at $3.1, 7.1$\,Hz for the first and second QPOs, respectively). As done in {\thisgrb} one of the power-laws was used to fit the low frequency red noise. 
%All these components are shown as dotted lines in Fig.\ref{timing2}.

The best PDS fit obtained had
reduced ${\chi}^2$ of $0.9,1.2$ for $19,8$ degrees of freedom for the {\it Fermi} and ASIM observations, respectively. The QPOs frequencies were measured to be $3.1,7.1$\,Hz for both QPOs, respectively. This means that the frequency of the QPOs was the same for {\it Fermi} and ASIM observations respectively (during the same time period). The ratio between these QPOs frequencies is also very similar to the one measured from {\thisgrb} (i.e. ${\approx}2$). The significance (without and with trials \footnote{We consider for the trials the number of frequency bins in the PDS.}) of these peaks is high, i.e. $8,5\,{\sigma}$ and $8,8\,{\sigma}$ (ASIM/LED) and $4,3\,{\sigma}$ and $5,4\,{\sigma}$ ({\it Fermi}) for the first and second QPOs, respectively. The characteristics of the QPOs observed in the
PDS of the observations are listed in Table \ref{ttiming2}.

%\vspace{-0.3cm}

  %\vspace{0.3cm}

\subsection{{\it Fermi} and ASIM timing analysis of {\thisgrboa} }

As done previously to obtain the exact value of the peak frequency of these QPO features, the PDS was fit
with three/four continuum components, i.e. two power-laws, and one/two Lorentzians (one for each QPO for {\it Fermi}/ASIM, respectively). The Lorentzian components were used to fit the QPO features (at $2.30{\pm}0.03,2.0^{+0.5}_{-0.10}$\,Hz for the ASIM and {\it Fermi}, respectively). The component fitted by a Lorentzian in the ASIM PDS with a centroid at $0.81{\pm}0.15$\,Hz is attributed to low-frequency noise. This component is much broader ($0.37{\pm}0.15$\,Hz) than the QPO ($(1-3){\times}10^{-3}$\,Hz) in both {\it Fermi} and ASIM datasets. 
%All these components are shown as dotted lines in Fig.\ref{timing3}.

The best PDS fit obtained had
reduced ${\chi}^2$ of $41.8,43.6$ for $40,32$ degrees of freedom for the {\it Fermi} and ASIM observations, respectively. The QPOs frequencies were measured to be $2.30{\pm}0.03,2.0^{+0.5}_{-0.10}$\,Hz for for {\it Fermi} and ASIM, respectively \footnote{We note that for \thisgrboa the time periods in which the QPO at ${\approx}2$\,Hz was found are not simultaneous but complementary.}. These QPOs frequencies are also very similar to the one measured from {\thisgrb} and {\thisgrbo} (i.e. $2.5-4,3$\,Hz), respectively. The significance of these peaks is high, i.e. (single trial) significance of $8\,{\sigma}$ (ASIM/HED) and $5\,{\sigma}$ ({\it Fermi}). The quality factor of the QPO was of
${\rm Q}=92,2000$ for the {\it Fermi} and ASIM datasets, respectively. The characteristics of the QPOs observed in the
PDS of the observations are listed in Table \ref{ttiming3}.

  %\vspace{0.3cm}

\subsection{{\it Fermi} and ASIM timing analysis of {\thisgrbob} }

The PDSs of both {\it Fermi} and ASIM datasets showed the same kind of noise in the form of three low-frequency QPOs. To obtain the exact value of the peak frequency of these QPO features, the PDS was fit
with five continuum components, i.e. two power-laws, and three Lorentzians (one for each QPO) in both {\it Fermi} and ASIM datasets. The Lorentzian components were used to fit the QPO features (at $22.1^{+1.9}_{-1.4},40.3^{+0.8}_{-3},59.3^{+1.5}_{-0.8}$\,Hz and $11.9{\pm}0.5,21.8{\pm}0.3,34.6{\pm}0.7$\,Hz for the ASIM and {\it Fermi}, respectively). There is an additional component fitted by a powerlaw at low-frequencies that we attribute to low-frequency noise. This component has is a steep powerlaw (${\Gamma}=2$) in both {\it Fermi} and ASIM datasets. 
%All these components are shown as dotted lines in Fig.\ref{timing4}.

The best PDS fit obtained had
reduced ${\chi}^2$ of $10.5,19.7$ for $7,23$ degrees of freedom for the {\it Fermi} and ASIM observations, respectively. The QPOs frequencies were measured to be $22.1^{+1.9}_{-1.4},40.3^{+0.8}_{-3},59.3^{+1.5}_{-0.8}$\,Hz and $11.9{\pm}0.5,21.8{\pm}0.3,34.6{\pm}0.7$\,Hz for ASIM and {\it Fermi}, respectively. This means that the frequency of the QPOs was not the same for {\it Fermi} and ASIM observations (only one of the three coincided during the same time period, i.e. the one at ${\approx}20\,Hz$) and might be due to the different energy ranges covered by both instruments. The QPO frequency at ${\approx}20$\,Hz is very similar to the one measured from {\thisgrb} by \cite{2022arXiv220502186X,2024ApJ...967...26C}. The significance of this peak is high ($8,8\,{\sigma}$ for single and taking into account trials, respectively) for both ASIM/HED and {\it Fermi}. The quality factor was of
${\rm Q}=5,12,10$ ({\it Fermi}) and ${\rm Q}=5,22,20$ (ASIM) for the first to third QPOs, respectively. The characteristics of the QPOs observed in the
PDS of the observations are listed in Table \ref{ttiming4}.

\subsection{The Bayesian method}

For estimating significances of the observed QPO peaks we used a Bayesian approach as proposed in \citet{2010MNRAS.402..307V} (see also \citealt{2013ApJ...768...87H}). In our case we fitted Lorentzians complementing the standard method by \citet{2010MNRAS.402..307V}.

The peak significance was obtained from the posterior predictive p-values that are the tail area probability of the Bayesian analogue of the Likelihood ratio test (LRT; \citealt{2013ApJ...768...87H}). The LRT statistic assesses the improvement $T_{LRT} = D_{min}(H0) - D_{min}(H1)$ that a more complicated model H1 (alternative hypothesis) gives with respect to its simpler version H0 (null-hypothesis), where D is twice the minus log likelihood (eq. 17 in \citealt{2010MNRAS.402..307V}). In the case of two-QPO power spectra and testing for the significance of the low frequency peak, the H1 model was ``continuum $+$ two Lorentzians" while H0 was ``continuum $+$ a Lorentzian" placed at the position of the high frequency peak. For the continuum we used the same model as above. Fitting both models to the observed power spectrum, we obtained ``the observed" statistic ${\rm T}_{\rm LRT}^{\rm obs}$, which was compared to the values measured from simulated periodograms (${\rm T}_{\rm LRT}^{\rm rep}$) in order to derive the p-value. The data-points of the latter were obtained from the values of the parameters sampled from the posterior distributions to the H0 model and adding $\chi^2$ distributed noise to the resultant smooth model periodogram. A number of 1000 spectra were used for each model and each of these realizations was fitted 100 times with a randomized start point.

In all the cases we analyzed unbinned priodograms and used the ``whittle" statistic in 
{\tt XSPEC} \footnote{\url{https://heasarc.gsfc.nasa.gov/xanadu/xspec/manual/XSappendixStatistics.html}}. For the Bayesian calculation, we 
employed the ``bxa" 
python package which uses the Nested Sampling integration algorithm as a Bayesian engine \citep{2014A&A...564A.125B,2016S&C....26..383B}. At first we assigned non-informative (log-uniform for normalizations and uniform for the rest parameters) priors to all the model parameters, but we found that Lorentzian components in many cases either fit high-frequency noise or became too wide to fit low-frequency continuum. In order to solve this issue, we redefined the Lorentzian model to took the quality factor (Q) instead of the width in Hz, and restricted it to be higher than 2 according to the definition of QPO \citep{1989ARA&A..27..517V}. To avoid moving the Lorentzian components to higher frequencies, we changed the position parameter (${\nu}$) from uniform to Gaussian priors, with mean and sigma values taken from the standard fitting (i.e. simple minimization of the fit statistic) in {\tt XSPEC}. The results of this procedure are shown in Tab. \ref{ttiming1}-\ref{ttiming4}.

%${\rm Q}={\nu}/{\rm W}$

\begin{table*}
\caption{{\bf Power Density Spectra (PDS) timing analysis results} of the light curve from \thisgrb using the \sw{powerlaw+lorentzian+lorentzian+powerlaw} function for {\it Fermi} (upper) and ASIM (lower) data. (T$_{\rm start}$ and T$_{\rm stop}$ have been referred with respect to the {\it Fermi} reference time (T$_{\rm 0, FERMI}$). The errors given are $1{\sigma}$.}
\label{ttiming1}
\begin{scriptsize}
\begin{center}
\begin{tabular}{|cc|cccccccccc|c|} \hline
T$_{\rm start}$  & T$_{\rm stop}$   & \boldmath $\it \Gamma_{\rm 0}$ & \boldmath $\rm N_{\rm 0}$ & \boldmath $\it \nu_{\rm QPO}$ & \boldmath $\rm FWHM_{\rm QPO}$ & \boldmath $\rm N_{\rm QPO}$  & \boldmath $\it \nu_{\rm v}$ & \boldmath $\rm FWHM_{\rm v}$ & \boldmath $\rm N_{\rm v}$  &  \boldmath $\it \Gamma_{\rm P}$  &  \boldmath $\rm N_{\rm P}$  & \bf $\chi^2$  (d.o.f.)   \\ 
(s) &  (s)  &  &  & ${\rm (Hz)}$ & ${\rm (Hz)}$ &   & ${\rm (Hz)}$ & ${\rm (Hz)}$ &   &    &    &    \\ \hline
5.8 & 8.8 & $1.29_{-0.23}^{+0.14}$   & $200{\pm}100$   & \boldmath $2.8{\pm}0.3$  & $(5_{-3}^{+200}){\rm E}-3$  &  $500{\pm}300$  & $5.2_{-0.8}^{+0.5}$    & $0.9_{-0.7}^{+0.9}$   & $370{\pm}170$   &  0\,(f) & $1.60{\pm}0.10$    &   20\,(40)        \\ 
QPO S/N (w. trials, Bayes.) &  &    &    & ($5{\sigma},2{\sigma}$)  &   &    & ($4{\sigma},2{\sigma}$)    &    &    &   &     &           \\ \hline
5.8 & 8.8 & $1.0{\pm}0.3$   & $60{\pm}30$    & \boldmath $2.9{\pm}0.4$  & $0.25_{-0.24}^{+0.9}$  & $240{\pm}140$   &   $5.2{\pm}1.0$ &   $0.013_{-0.010}^{+0.90}$  &  $110{\pm}60$  &  0\,(f)   &  $1.68{\pm}0.15$  &      32\,(40)        \\ 
QPO S/N (w. trials, Bayes.) &  &    &     & ($5{\sigma},1{\sigma}$)  &   &    &   ($4{\sigma},1{\sigma}$) &     &    &     &    &              \\ \hline
\end{tabular}
\end{center}
\end{scriptsize}
\end{table*}

  \vspace{-0.3cm}

\begin{table*}
\caption{{\bf Power Density Spectra (PDS) timing analysis results} of the light curve from GRB-220910A using the \sw{powerlaw+lorentzian+lorentzian+powerlaw} function for {\it Fermi} (upper) and ASIM/LED (lower). }
%(T$_{\rm start}$ and T$_{\rm stop}$ have been referred with respect to the {\it Fermi} reference time (T$_{\rm 0, FERMI}$). The errors given are $1{\sigma}$.}
\label{ttiming2}
\begin{scriptsize}
\begin{center}
\begin{tabular}{|cc|cccccccccc|c|} \hline
T$_{\rm start}$ (s) & T$_{\rm stop}$ (s)  & \boldmath $\it \Gamma_{\rm 0}$ & \boldmath $\rm N_{\rm 0}$ & \boldmath $\it \nu_{\rm QPO}$ & \boldmath $\rm FWHM_{\rm QPO}$ & \boldmath $\rm N_{\rm QPO}$  & \boldmath $\it \nu_{\rm v}$ & \boldmath $\rm FWHM_{\rm v}$ & \boldmath $\rm N_{\rm v}$  &  \boldmath $\it \Gamma_{\rm P}$  &  \boldmath $\rm N_{\rm P}$  & \bf $\chi^2$  (d.o.f.)   \\ 
(s) & (s)  &  &  & ${\rm (Hz)}$ & ${\rm (Hz)}$ &   & ${\rm (Hz)}$ & ${\rm (Hz)}$ &   &    &    &    \\ \hline
 7.3  & 10.0 & $5{\pm}4$   & $2.5{\rm E}3_{-2.2{\rm E}3}^{+2{\rm E}4}$    & \boldmath $3.1_{-0.2}^{+0.6}$  & $0.5_{-0.3}^{+0.8}$  &  $110_{-50}^{+60}$  & $7.1{\pm}0.3$   & $0.6{\pm}0.5$  & $50{\pm}20$   &  0\,(f) & $1.67{\pm}0.20$    &   18\,(19)        \\ 
QPO S/N (w. trials, Bayes.)   &  &    &     & ($4{\sigma},1{\sigma}$)  &   &    & ($5{\sigma},2.4{\sigma}$)   &   &    &   &     &           \\ \hline
 7.3  & 10.0 & $10{\pm}9$  & $5{\rm E}4{\pm}4{\rm E}4$  & \boldmath $3.1_{-0.3}^{+0.4}$  & $0.8_{-0.5}^{+0.4}$  &  $600{\pm}300$   &   $7.1{\pm}0.3$ &   $0.9{\pm}0.6$  &  $360_{-120}^{+110} $  & 0\,(f)     &  $1.5{\pm}0.4$  &        10\,(8)        \\ \
QPO S/N (w. trials, Bayes.)   &  &   &   & ($8{\sigma},2{\sigma}$)  &   &     &   ($8{\sigma},2{\sigma}$) &     &     &      &    &               \\ \hline
\end{tabular}
\end{center}
\end{scriptsize}
\end{table*}

  \vspace{-0.3cm}

\begin{table*}
\caption{{\bf Power Density Spectra (PDS) timing analysis results} of the light curve from \thisgrboa using the \sw{powerlaw+lorentzian+powerlaw} and \sw{lorentzian+lorentzian+lorentzian+powerlaw} functions for {\it Fermi} (upper) and ASIM (lower), respectively. }
%(T$_{\rm start}$ and T$_{\rm stop}$ have been referred with respect to the {\it Fermi} reference time (T$_{\rm 0, FERMI}$). The errors given are $1{\sigma}$.}
\label{ttiming3}
\begin{scriptsize}
\begin{center}
\begin{tabular}{|cc|cccccccccc|c|} \hline
 T$_{\rm start}$  & T$_{\rm stop}$   & \boldmath $\it \Gamma_{\rm 0}$ & \boldmath $\rm N_{\rm 0}$ & $-$ & \boldmath $\it \nu_{\rm QPO}$ & \boldmath $\rm FWHM_{\rm QPO}$ & \boldmath $\rm N_{\rm QPO}$  & \boldmath $\it \nu_{\rm v}$ & \boldmath $\rm FWHM_{\rm v}$ & \boldmath $\rm N_{\rm v}$  &    \boldmath $\rm N_{\rm P}$  & \bf $\chi^2$  (d.o.f.)   \\ 
  (s) & (s)  &  &  & $-$ & ${\rm (Hz)}$ & ${\rm (Hz)}$ &   & ${\rm (Hz)}$ & ${\rm (Hz)}$ &   &      &    \\ \hline
-9.0 & 13.0 & $1.7{\pm}0.4$   & $7.3{\pm}1.6$     & $-$ & \boldmath $2.30_{-0.03}^{+0.01}$  & $0.025_{-0.015}^{+0.10}$  &  $8{\pm}3$  & $-$    & $-$   & $-$   & $1.89{\pm}0.08$       &  41.8\,(40)        \\
QPO S/N (w. trials, Bayes.) &  &    &      &  & ($5{\sigma},{\ge}3.3{\sigma}$)  &   &    &    &   &    &        &         \\ \hline
T$_{\rm start}$  & T$_{\rm stop}$   & \boldmath $\it \nu_{\rm 0}$ & \boldmath $\rm FWHM_{\rm 0}$ & \boldmath $\rm N_{\rm 0}$ & \boldmath $\it \nu_{\rm QPO}$ & \boldmath $\rm FWHM_{\rm QPO}$ & \boldmath $\rm N_{\rm QPO}$  & \boldmath $\it \nu_{\rm v}$ & \boldmath $\rm FWHM_{\rm v}$ & \boldmath $\rm N_{\rm v}$  &    \boldmath $\rm N_{\rm P}$  & \bf $\chi^2$  (d.o.f.)   \\ 
(s)  & (s)   & (Hz) & (Hz) &  & ${\rm (Hz)}$ & ${\rm (Hz)}$ &   & ${\rm (Hz)}$ & ${\rm (Hz)}$ &   &      &   \\ \hline
13 & 30 & $0.81{\pm}0.15$   & $0.37{\pm}0.15$     & $700{\pm}200$ & \boldmath $2.0_{-0.10}^{+0.5}$  & $(8_{-1.0}^{+270}){\times}10^{-4}$  &  $130{\pm}60$  & $3.5_{-0.7}^{+0.9}$    & $8.2{\pm}1.8$   & $210{\pm}40$    &  $1.3{\pm}0.3$      &43.6\,(32)        \\ 
QPO S/N (w. trials, Bayes.) &  &    &      &  & ($8{\sigma},2{\sigma}$)  &   &    & ($4{\sigma},2{\sigma}$)    &    &    &       &        \\ \hline
\end{tabular}
\end{center}
\end{scriptsize}
\end{table*}

  \vspace{-0.3cm}

\begin{table*}
\caption{{\bf Power Density Spectra (PDS) timing analysis results} of the light curve from \thisgrbob using the \sw{powerlaw+lorentzian+lorentzian+lorentzian+powerlaw} functions for {\it Fermi} (upper) and ASIM (lower). }
%( T$_{\rm start}$ and T$_{\rm stop}$ have been referred with respect to the {\it Fermi} reference time (T$_{\rm 0, FERMI}$). The errors given are $1{\sigma}$.}
\label{ttiming4}
\begin{scriptsize}
\begin{center}
\begin{tabular}{|cc|cccccccccc|c|} \hline
 T$_{\rm start}$  & T$_{\rm stop}$   & \boldmath $\it \nu_{\rm 3}$ & \boldmath $\rm FWHM_{\rm 3}$ & \boldmath $\rm N_{\rm 3}$ & \boldmath $\it \nu_{\rm 2}$ & \boldmath $\rm FWHM_{\rm 2}$ & \boldmath $\rm N_{\rm 2}$  & \boldmath $\it \nu_{\rm 1}$ & \boldmath $\rm FWHM_{\rm 1}$ & \boldmath $\rm N_{\rm 1}$  &   \boldmath $\rm N_{\rm P}$  & \bf $\chi^2$  (d.o.f.)   \\ 
 (s) & (s)  & ${\rm (Hz)}$ & ${\rm (Hz)}$ &  & ${\rm (Hz)}$ & ${\rm (Hz)}$ &   & ${\rm (Hz)}$ & ${\rm (Hz)}$ &   &     &    \\ \hline
 0.15 & 5.0 & $34.6{\pm}0.7$   & $3.5_{-1.4}^{+2.2}$     & $60{\pm}20$ & \boldmath $21.8{\pm}0.3$  & $1.8_{-0.3}^{+1.1}$  &  $250_{-50}^{+40}$  &  $11.9{\pm}0.5$    & $2.4{\pm}0.7$   & $370{\pm}80$   &   $1.7{\pm}0.8$      &  10.5\,(7)    \\ 
QPO S/N (w. trials, Bayes.)   &   & ($2.5{\sigma},1{\sigma}$)   &      &  & ($8{\sigma},3{\sigma}$)  &   &    &  ($5{\sigma},1.5{\sigma}$)    &    &    &        &      \\ \hline
T$_{\rm start}$  & T$_{\rm stop}$   & \boldmath $\it \nu_{\rm 3}$ & \boldmath $\rm FWHM_{\rm 3}$ & \boldmath $\rm N_{\rm 3}$ & \boldmath $\it \nu_{\rm 2}$ & \boldmath $\rm FWHM_{\rm 2}$ & \boldmath $\rm N_{\rm 2}$  & \boldmath $\it \nu_{\rm 1}$ & \boldmath $\rm FWHM_{\rm 1}$ & \boldmath $\rm N_{\rm 1}$ & \boldmath $\rm N_{\rm P}$  & \bf $\chi^2$  (d.o.f.)   \\ 
 (s) &  (s)  & ${\rm (Hz)}$ & ${\rm (Hz)}$ &  & ${\rm (Hz)}$ & ${\rm (Hz)}$ &   & ${\rm (Hz)}$ & ${\rm (Hz)}$ &  &   &    \\ \hline
0.15 & 1.0 & $59.3_{-0.8}^{+1.5}$   & $2.8_{-2.5}^{+4}$     & $280{\pm}110$ &  $40.3_{-3}^{+0.8}$  & $1.8_{-1.7}^{+6}$  &  $240_{-110}^{+160}$  & \boldmath $22.1_{-1.4}^{+1.9}$    & $4.7_{-2.0}^{2.8}$   & $1070{\pm}400$   &    $1.5{\pm}0.3$      &    19.7\,(23)    \\ 
QPO S/N (w. trials, Bayes.) &  & ($4.5{\sigma},2{\sigma}$)   &      &  &  ($4{\sigma},1{\sigma}$)  &   &    & ($8{\sigma},2{\sigma}$)    &    &    &          &        \\ \hline
\end{tabular}
\end{center}
\end{scriptsize}
\end{table*}

%  \vspace{-0.3cm}

%\begin{table}
%\caption{{\bf Minimum variability time scales of GRBs in the sample extracted from Fermi/GBM NaI light curves.}}
%\begin{scriptsize}
%\begin{center}
%\begin{tabular}{ccc} \hline
% GRB Name & QPO range & Pre-QPO range  \\
%\hline
%	       & sec    & sec  \\ 
%\hline
%GRB 180720B & 0.026 $\pm$ 0.003 & 0.093 $\pm$ 0.030 \\
%GRB 181222B & 0.007 $\pm$ 0.001 & 0.018 $\pm$ 0.007 \\
%GRB 211211A & 0.022 $\pm$ 0.007 & 0.028 $\pm$ 0.009 \\
%GRB 220910A & 0.025 $\pm$ 0.005 & 0.034 $\pm$ 0.010 \\
%\hline
%\end{tabular}
%\end{center}
%\end{scriptsize}
%\label{mtstab}
%\end{table}

%\vspace{0.6cm}

%  \clearpage
%\newpage

\section{Discussion and Conclusions}\label{discussion} 

In this paper we analyzed the first seconds of the ASIM and {\it Fermi} data for GRB~180720B, GRB~181222B, GRB~211211A and GRB~220910A that correspond to the first period of their activity 
since the {\it Fermi} trigger time. All of them are lGRBs with the exception of \thisgrbob that is a sGRB (Fig. \ref{mvts}). As derived from the timing analysis we infer the presence of a low-frequency Quasi-Periodic Oscillation (QPO) 
with a frequency at $2.5-3$\,Hz in all these GRBs. The significance of the low-frequency QPO signal is high (i.e. significance ${\gtrsim}5{\sigma}$) from both {\it Fermi} and ASIM periodograms of our observations. This low-frequency QPO could be identified as the orbital period of the binary merger (NS-BH or NS-NS) at the previous stage before its collapse into a single compact object (BH or NS) as previously proposed (\citealt{2022A&A...664A.177S}; see their Tab. 1). Due to the short orbital period of the merger (0.3-0.4\,s) the system would be tidally locked (and synchronized) and this period would correspond to the spin of the NS component (a magnetar). The NS could produce the (also) observed 
QPOs at ${\approx}20$\,Hz in \thisgrb and \thisgrbob through the star-quakes followed by crustal vibrations occurring on its surface at the previous moments to the coalescence/collapse of the merger. In the case of the kilonova (GRB~211211A) that
produced the high-energy emission in the form of a gamma-ray burst \citep{2022Natur.612..228T} the nature of the binary components is under active discussion. We consider that our findings support that the merger nature is very likely in GRB~211211A and the other GRB similar events.
%Similar instances of such events occur often in the so-called magnetars 
%(i.e. magnetized and highly spinning isolated NSs) 
%but never observed before in GRBs. 

The $2.5-3$\,Hz QPO reported in this paper is of low frequency and its value could be the first detection of a low-frequency feature compatible with the spin of the NS (magnetar) in a binary merger (i.e. which would be equal to the orbital period of the binary merger) in these GRBs before coalescence (compatible with the scenario proposed by \citealt{2022A&A...664A.177S}). Our findings would be in line of the fact that compact object ``mergers” may be a non-negligible fraction of the lGRB population (as suggested by \citealt{2024MNRAS.tmp.2439L}).  

The low-frequency QPO is compatible with the higher frequency QPO at ${\approx}20$\,Hz detected previously by \citet{2022arXiv220502186X,2024ApJ...967...26C} in the modulated {\it Fermi} (8-200)\,keV X-ray emission from GRB~211211A. The latter
%\footnote{They detected the QPO at ${\approx}20$\,Hz in a different time period than the one used in our study.}
would correspond to the node-less, torsional modes of the magnetized NS long before the merger occurs (even though global or discrete Alfv\'en modes are also viable explanations). Indeed we detect a significant ${\approx}20$\,Hz QPO (together with its less-significant harmonics) in the case of the (short duration) GRB~181222B. We do not detect the low-frequency $2.5-3$\,Hz QPO in GRB~181222B because of its extremely short duration (${\approx}0.3$\,s) but we detect clearly the ${\approx}20$\,Hz QPO (and probably its harmonics).

Even though rare the current work does not constitute the first claim into the presence of low-frequency QPOs in GRBs neither in magnetars. High frequency QPOs
are known to happen in giant flares from magnetars at the frequency range of $500-2000$\,Hz \citep{2021Natur.600..621C}. At lower frequencies there are also claims of lower frequency QPOs occurring during giant flares as well ($150$\,Hz; \citealt{2024MNRAS.527..855L}). In all these cases torsional oscillations caused by star-quakes could explain the QPOs observed. \citet{2024ApJ...973..126Z} report the finding of a low-frequency QPO in the GRB 210514A (with a $3{\sigma}$ confidence) corresponding to a period of 11\,s. They associate it as the precession frequency of an accretion disc around a single magnetar before its collapse as a BH. This is opposite to our claims that a companion is needed to produce the GRB explosion. 

\citet{2024ApJ...973L..33C} is in line of our claims of a detection of low-frequency Quasi-Periodic Modulations (QPM) at a frequency of ${\le}1$\,Hz in a sample of GRBs (GRB 230307A, GRB 060614 snd 211211A). They claim that in light of recent claims of WD-NS and/or WD-BH mergers in the literature \citep{2024ApJ...964L...9W} the 
QPM would be originated by the precession of the jet due to the non-negligible residual orbital eccentricity of the WD-NS/BH merger. This scenario is different to ours, in which the super-flare originated by the crust in the NS (necessarily a magnetar) would be the main responsible for the X-ray and gamma-ray emission and modulation observed in the light curves of these GRBs.

\clearpage
\newpage

\section{Acknowledgements}
MCG acknowledges S. Xiao, M. Marisaldi and A. Mezentsev for useful discussions and feedback. MCG acknowledges financial support from the Spanish Ministry project MCI/AEI/PID2023-149817OB-C31. RSR, YDH, MCG, SG, IPG, EJFG, RGB, MPV, and AJCT acknowledge financial support from the Spanish Ministry project PID2023-151905OB-I00 and the Severo Ochoa grant CEX2021-001131-S funded by MICIU/AEI/10.13039/501100011033. RG and SBP acknowledge BRICS grant {DST/IMRCD/BRICS /PilotCall1 /ProFCheap/2017(G)} for the financial support. RG and SBP also acknowledge the financial support of ISRO under AstroSat archival Data utilization program (DS$\_$2B-13013(2)/1/2021-Sec.2). RG was sponsored by the National Aeronautics and Space Administration (NASA) through a contract with ORAU. AJCT acknowledges support from the Junta de Andalucia Project P20$\_$01068. MM, AL, AM, and N{\O} acknowledge financial support from the Research Council of Norway under Contracts 208028/F50 and 223252/F50 (CoE). This research has used data obtained through the HEASARC Online Service, provided by the NASA-GSFC, in support of NASA High Energy Astrophysics Programs. The NASA’s FGST ({\it Fermi}) mission is an astrophysics and particle physics partnership, developed in collaboration with the U.S. Department of Energy, along with important contributions from academic institutions and partners in France, Germany, Italy, Japan, Sweden, and the United States. ASIM is a mission of ESA’s SciSpace programme for scientific utilization of the {\it ISS} and non-{\it ISS} space exploration platforms and space environment analogues. ASIM and the ASIM Science Data Centre are funded by ESA and by national grants of Denmark, Norway and Spain.

%%%%%%%%%%%%%%%%%%%%%%%%%%%%%%%%%%%%%%%%%%%%%%%%%%
\section*{Data Availability}

The {\it Fermi} data was obtained through the website \url{https://heasarc.gsfc.nasa.gov/FTP/fermi/data/gbm/bursts/}. The ASIM \citep{2019arXiv190612178N} data presented in this work can be made available based on the individual request to the corresponding authors. The Power Density Spectra (PDS) from the source light curves were analyzed using the {\tt FTOOL}-{\tt powspec} from the {\tt XRONOS} package of timing tools \footnote{\url{https://heasarc.gsfc.nasa.gov/docs/xanadu/xronos/xronos.html }}. Fits were performed with the {\tt XSPEC} package \footnote{\url{https://heasarc.gsfc.nasa.gov/xanadu/xspec/}}.

\bibliographystyle{mnras}
\bibliography{GRB211211A}

\begin{thebibliography}{}
\makeatletter
\relax
\def\mn@urlcharsother{\let\do\@makeother \do\$\do\&\do\#\do\^\do\_\do\%\do\~}
\def\mn@doi{\begingroup\mn@urlcharsother \@ifnextchar [ {\mn@doi@}
  {\mn@doi@[]}}
\def\mn@doi@[#1]#2{\def\@tempa{#1}\ifx\@tempa\@empty \href
  {http://dx.doi.org/#2} {doi:#2}\else \href {http://dx.doi.org/#2} {#1}\fi
  \endgroup}
\def\mn@eprint#1#2{\mn@eprint@#1:#2::\@nil}
\def\mn@eprint@arXiv#1{\href {http://arxiv.org/abs/#1} {{\tt arXiv:#1}}}
\def\mn@eprint@dblp#1{\href {http://dblp.uni-trier.de/rec/bibtex/#1.xml}
  {dblp:#1}}
\def\mn@eprint@#1:#2:#3:#4\@nil{\def\@tempa {#1}\def\@tempb {#2}\def\@tempc
  {#3}\ifx \@tempc \@empty \let \@tempc \@tempb \let \@tempb \@tempa \fi \ifx
  \@tempb \@empty \def\@tempb {arXiv}\fi \@ifundefined
  {mn@eprint@\@tempb}{\@tempb:\@tempc}{\expandafter \expandafter \csname
  mn@eprint@\@tempb\endcsname \expandafter{\@tempc}}}

\bibitem[\protect\citeauthoryear{{Abbott} et~al.,}{{Abbott}
  et~al.}{2017}]{2017ApJ...848L..12A}
{Abbott} B.~P.,  et~al., 2017, \mn@doi [\apjl] {10.3847/2041-8213/aa91c9},
  \href {https://ui.adsabs.harvard.edu/abs/2017ApJ...848L..12A} {848, L12}

\bibitem[\protect\citeauthoryear{{Belloni} \& {Hasinger}}{{Belloni} \&
  {Hasinger}}{1990}]{1990A&A...230..103B}
{Belloni} T.,  {Hasinger} G.,  1990, \aap, \href
  {https://ui.adsabs.harvard.edu/abs/1990A&A...230..103B} {230, 103}

\bibitem[\protect\citeauthoryear{{Buchner}}{{Buchner}}{2016}]{2016S&C....26..383B}
{Buchner} J.,  2016, \mn@doi [Statistics and Computing]
  {10.1007/s11222-014-9512-y}, \href
  {https://ui.adsabs.harvard.edu/abs/2016S&C....26..383B} {26, 383}

\bibitem[\protect\citeauthoryear{{Buchner} et~al.,}{{Buchner}
  et~al.}{2014}]{2014A&A...564A.125B}
{Buchner} J.,  et~al., 2014, \mn@doi [\aap] {10.1051/0004-6361/201322971},
  \href {https://ui.adsabs.harvard.edu/abs/2014A&A...564A.125B} {564, A125}

\bibitem[\protect\citeauthoryear{{Camisasca} et~al.,}{{Camisasca}
  et~al.}{2023}]{2023A&A...671A.112C}
{Camisasca} A.~E.,  et~al., 2023, \mn@doi [\aap] {10.1051/0004-6361/202245657},
  \href {https://ui.adsabs.harvard.edu/abs/2023A&A...671A.112C} {671, A112}

\bibitem[\protect\citeauthoryear{{Castro-Tirado} et~al.,}{{Castro-Tirado}
  et~al.}{2021}]{2021Natur.600..621C}
{Castro-Tirado} A.~J.,  et~al., 2021, \mn@doi [\nat]
  {10.1038/s41586-021-04101-1}, \href
  {https://ui.adsabs.harvard.edu/abs/2021Natur.600..621C} {600, 621}

\bibitem[\protect\citeauthoryear{{Chen}, {Shen}, {Tan}, {Wang}, {Xiong}, {Chen}
   \& {Zhang}}{{Chen} et~al.}{2024}]{2024ApJ...973L..33C}
{Chen} J.,  {Shen} R.-F.,  {Tan} W.-J.,  {Wang} C.-W.,  {Xiong} S.-L.,  {Chen}
  R.-C.,   {Zhang} B.-B.,  2024, \mn@doi [\apjl] {10.3847/2041-8213/ad7737},
  \href {https://ui.adsabs.harvard.edu/abs/2024ApJ...973L..33C} {973, L33}

\bibitem[\protect\citeauthoryear{{Chirenti}, {Dichiara}, {Lien}  \&
  {Miller}}{{Chirenti} et~al.}{2024}]{2024ApJ...967...26C}
{Chirenti} C.,  {Dichiara} S.,  {Lien} A.,   {Miller} M.~C.,  2024, \mn@doi
  [\apj] {10.3847/1538-4357/ad3bb7}, \href
  {https://ui.adsabs.harvard.edu/abs/2024ApJ...967...26C} {967, 26}

\bibitem[\protect\citeauthoryear{{Gao}, {Lei}  \& {Zhu}}{{Gao}
  et~al.}{2022}]{2022ApJ...934L..12G}
{Gao} H.,  {Lei} W.-H.,   {Zhu} Z.-P.,  2022, \mn@doi [\apjl]
  {10.3847/2041-8213/ac80c7}, \href
  {https://ui.adsabs.harvard.edu/abs/2022ApJ...934L..12G} {934, L12}

\bibitem[\protect\citeauthoryear{{Gehrels} et~al.,}{{Gehrels}
  et~al.}{2006}]{2006Natur.444.1044G}
{Gehrels} N.,  et~al., 2006, \mn@doi [\nat] {10.1038/nature05376}, \href
  {https://ui.adsabs.harvard.edu/abs/2006Natur.444.1044G} {444, 1044}

\bibitem[\protect\citeauthoryear{{Golkhou} \& {Butler}}{{Golkhou} \&
  {Butler}}{2014}]{2014ApJ...787...90G}
{Golkhou} V.~Z.,  {Butler} N.~R.,  2014, \mn@doi [\apj]
  {10.1088/0004-637X/787/1/90}, \href
  {https://ui.adsabs.harvard.edu/abs/2014ApJ...787...90G} {787, 90}

\bibitem[\protect\citeauthoryear{{Golkhou}, {Butler}  \&
  {Littlejohns}}{{Golkhou} et~al.}{2015}]{2015ApJ...811...93G}
{Golkhou} V.~Z.,  {Butler} N.~R.,   {Littlejohns} O.~M.,  2015, \mn@doi [\apj]
  {10.1088/0004-637X/811/2/93}, \href
  {https://ui.adsabs.harvard.edu/abs/2015ApJ...811...93G} {811, 93}

\bibitem[\protect\citeauthoryear{{Hakkila}, {Giblin}, {Norris}, {Fragile}  \&
  {Bonnell}}{{Hakkila} et~al.}{2008}]{2008ApJ...677L..81H}
{Hakkila} J.,  {Giblin} T.~W.,  {Norris} J.~P.,  {Fragile} P.~C.,   {Bonnell}
  J.~T.,  2008, \mn@doi [\apjl] {10.1086/588094}, \href
  {https://ui.adsabs.harvard.edu/abs/2008ApJ...677L..81H} {677, L81}

\bibitem[\protect\citeauthoryear{{Huppenkothen} et~al.,}{{Huppenkothen}
  et~al.}{2013}]{2013ApJ...768...87H}
{Huppenkothen} D.,  et~al., 2013, \mn@doi [\apj] {10.1088/0004-637X/768/1/87},
  \href {https://ui.adsabs.harvard.edu/abs/2013ApJ...768...87H} {768, 87}

\bibitem[\protect\citeauthoryear{{Leahy}, {Darbro}, {Elsner}, {Weisskopf},
  {Sutherland}, {Kahn}  \& {Grindlay}}{{Leahy}
  et~al.}{1983}]{1983ApJ...266..160L}
{Leahy} D.~A.,  {Darbro} W.,  {Elsner} R.~F.,  {Weisskopf} M.~C.,  {Sutherland}
  P.~G.,  {Kahn} S.,   {Grindlay} J.~E.,  1983, \mn@doi [\apj]
  {10.1086/160766}, \href
  {https://ui.adsabs.harvard.edu/abs/1983ApJ...266..160L} {266, 160}

\bibitem[\protect\citeauthoryear{{Li}, {Kang}, {Hu}, {Shao}, {Xia}  \&
  {Xu}}{{Li} et~al.}{2024}]{2024MNRAS.527..855L}
{Li} H.-B.,  {Kang} Y.,  {Hu} Z.,  {Shao} L.,  {Xia} C.-J.,   {Xu} R.-X.,
  2024, \mn@doi [\mnras] {10.1093/mnras/stad3204}, \href
  {https://ui.adsabs.harvard.edu/abs/2024MNRAS.527..855L} {527, 855}

\bibitem[\protect\citeauthoryear{{Liu} \& {Zou}}{{Liu} \&
  {Zou}}{2024}]{2024JCAP...07..070L}
{Liu} D.-J.,  {Zou} Y.-C.,  2024, \mn@doi [\jcap]
  {10.1088/1475-7516/2024/07/070}, \href
  {https://ui.adsabs.harvard.edu/abs/2024JCAP...07..070L} {2024, 070}

\bibitem[\protect\citeauthoryear{{Lloyd-Ronning}, {Johnson}, {Sanderbeck},
  {Silva}  \& {Cheng}}{{Lloyd-Ronning} et~al.}{2024}]{2024MNRAS.tmp.2439L}
{Lloyd-Ronning} N.~M.,  {Johnson} J.,  {Sanderbeck} P.~U.,  {Silva} M.,
  {Cheng} R.~M.,  2024, \mn@doi [\mnras] {10.1093/mnras/stae2502}, \href
  {https://ui.adsabs.harvard.edu/abs/2024MNRAS.tmp.2439L} {}

\bibitem[\protect\citeauthoryear{{L{\"u}} et~al.,}{{L{\"u}}
  et~al.}{2022}]{2022ApJ...931L..23L}
{L{\"u}} H.-J.,  et~al., 2022, \mn@doi [\apjl] {10.3847/2041-8213/ac6e3a},
  \href {https://ui.adsabs.harvard.edu/abs/2022ApJ...931L..23L} {931, L23}

\bibitem[\protect\citeauthoryear{{MacLachlan}, {Shenoy}, {Sonbas}, {Dhuga},
  {Eskandarian}, {Maximon}  \& {Parke}}{{MacLachlan}
  et~al.}{2012}]{2012MNRAS.425L..32M}
{MacLachlan} G.~A.,  {Shenoy} A.,  {Sonbas} E.,  {Dhuga} K.~S.,  {Eskandarian}
  A.,  {Maximon} L.~C.,   {Parke} W.~C.,  2012, \mn@doi [\mnras]
  {10.1111/j.1745-3933.2012.01295.x}, \href
  {https://ui.adsabs.harvard.edu/abs/2012MNRAS.425L..32M} {425, L32}

\bibitem[\protect\citeauthoryear{{MacLachlan} et~al.,}{{MacLachlan}
  et~al.}{2013a}]{2013MNRAS.432..857M}
{MacLachlan} G.~A.,  et~al., 2013a, \mn@doi [\mnras] {10.1093/mnras/stt241},
  \href {https://ui.adsabs.harvard.edu/abs/2013MNRAS.432..857M} {432, 857}

\bibitem[\protect\citeauthoryear{{MacLachlan}, {Shenoy}, {Sonbas}, {Coyne},
  {Dhuga}, {Eskandarian}, {Maximon}  \& {Parke}}{{MacLachlan}
  et~al.}{2013b}]{2013MNRAS.436.2907M}
{MacLachlan} G.~A.,  {Shenoy} A.,  {Sonbas} E.,  {Coyne} R.,  {Dhuga} K.~S.,
  {Eskandarian} A.,  {Maximon} L.~C.,   {Parke} W.~C.,  2013b, \mn@doi [\mnras]
  {10.1093/mnras/stt1701}, \href
  {https://ui.adsabs.harvard.edu/abs/2013MNRAS.436.2907M} {436, 2907}

\bibitem[\protect\citeauthoryear{{Neubert} et~al.,}{{Neubert}
  et~al.}{2019}]{2019arXiv190612178N}
{Neubert} T.,  et~al., 2019, arXiv e-prints, \href
  {https://ui.adsabs.harvard.edu/abs/2019arXiv190612178N} {p. arXiv:1906.12178}

\bibitem[\protect\citeauthoryear{{Norris}}{{Norris}}{2002}]{2002ApJ...579..386N}
{Norris} J.~P.,  2002, \mn@doi [\apj] {10.1086/342747}, \href
  {https://ui.adsabs.harvard.edu/abs/2002ApJ...579..386N} {579, 386}

\bibitem[\protect\citeauthoryear{{Pandey} et~al.,}{{Pandey}
  et~al.}{2019}]{2019MNRAS.485.5294P}
{Pandey} S.~B.,  et~al., 2019, \mn@doi [\mnras] {10.1093/mnras/stz530}, \href
  {https://ui.adsabs.harvard.edu/abs/2019MNRAS.485.5294P} {485, 5294}

\bibitem[\protect\citeauthoryear{{Rastinejad} et~al.,}{{Rastinejad}
  et~al.}{2022}]{2022arXiv220410864R}
{Rastinejad} J.~C.,  et~al., 2022, arXiv e-prints, \href
  {https://ui.adsabs.harvard.edu/abs/2022arXiv220410864R} {p. arXiv:2204.10864}

\bibitem[\protect\citeauthoryear{{Sonbas}, {MacLachlan}, {Dhuga}, {Veres},
  {Shenoy}  \& {Ukwatta}}{{Sonbas} et~al.}{2015}]{2015ApJ...805...86S}
{Sonbas} E.,  {MacLachlan} G.~A.,  {Dhuga} K.~S.,  {Veres} P.,  {Shenoy} A.,
  {Ukwatta} T.~N.,  2015, \mn@doi [\apj] {10.1088/0004-637X/805/2/86}, \href
  {https://ui.adsabs.harvard.edu/abs/2015ApJ...805...86S} {805, 86}

\bibitem[\protect\citeauthoryear{{Suvorov}, {Kuan}  \& {Kokkotas}}{{Suvorov}
  et~al.}{2022}]{2022A&A...664A.177S}
{Suvorov} A.~G.,  {Kuan} H.~J.,   {Kokkotas} K.~D.,  2022, \mn@doi [\aap]
  {10.1051/0004-6361/202244082}, \href
  {https://ui.adsabs.harvard.edu/abs/2022A&A...664A.177S} {664, A177}

\bibitem[\protect\citeauthoryear{{Tanvir}, {Levan}, {Fruchter}, {Hjorth},
  {Hounsell}, {Wiersema}  \& {Tunnicliffe}}{{Tanvir}
  et~al.}{2013}]{2013Natur.500..547T}
{Tanvir} N.~R.,  {Levan} A.~J.,  {Fruchter} A.~S.,  {Hjorth} J.,  {Hounsell}
  R.~A.,  {Wiersema} K.,   {Tunnicliffe} R.~L.,  2013, \mn@doi [\nat]
  {10.1038/nature12505}, \href
  {https://ui.adsabs.harvard.edu/abs/2013Natur.500..547T} {500, 547}

\bibitem[\protect\citeauthoryear{{Tarnopolski} \& {Marchenko}}{{Tarnopolski} \&
  {Marchenko}}{2021}]{2021ApJ...911...20T}
{Tarnopolski} M.,  {Marchenko} V.,  2021, \mn@doi [\apj]
  {10.3847/1538-4357/abe5b1}, \href
  {https://ui.adsabs.harvard.edu/abs/2021ApJ...911...20T} {911, 20}

\bibitem[\protect\citeauthoryear{{Troja} et~al.,}{{Troja}
  et~al.}{2019}]{2019MNRAS.489.2104T}
{Troja} E.,  et~al., 2019, \mn@doi [\mnras] {10.1093/mnras/stz2255}, \href
  {https://ui.adsabs.harvard.edu/abs/2019MNRAS.489.2104T} {489, 2104}

\bibitem[\protect\citeauthoryear{{Troja} et~al.,}{{Troja}
  et~al.}{2022}]{2022Natur.612..228T}
{Troja} E.,  et~al., 2022, \mn@doi [\nat] {10.1038/s41586-022-05327-3}, \href
  {https://ui.adsabs.harvard.edu/abs/2022Natur.612..228T} {612, 228}

\bibitem[\protect\citeauthoryear{{Ukwatta} et~al.,}{{Ukwatta}
  et~al.}{2010}]{2010ApJ...711.1073U}
{Ukwatta} T.~N.,  et~al., 2010, \mn@doi [\apj] {10.1088/0004-637X/711/2/1073},
  \href {https://ui.adsabs.harvard.edu/abs/2010ApJ...711.1073U} {711, 1073}

\bibitem[\protect\citeauthoryear{{Ukwatta} et~al.,}{{Ukwatta}
  et~al.}{2012}]{2012MNRAS.419..614U}
{Ukwatta} T.~N.,  et~al., 2012, \mn@doi [\mnras]
  {10.1111/j.1365-2966.2011.19723.x}, \href
  {https://ui.adsabs.harvard.edu/abs/2012MNRAS.419..614U} {419, 614}

\bibitem[\protect\citeauthoryear{{Vaughan}}{{Vaughan}}{2010}]{2010MNRAS.402..307V}
{Vaughan} S.,  2010, \mn@doi [\mnras] {10.1111/j.1365-2966.2009.15868.x}, \href
  {https://ui.adsabs.harvard.edu/abs/2010MNRAS.402..307V} {402, 307}

\bibitem[\protect\citeauthoryear{{Wang}, {Yu}, {Ren}, {Yang}, {Zou}  \&
  {Zhu}}{{Wang} et~al.}{2024}]{2024ApJ...964L...9W}
{Wang} X.~I.,  {Yu} Y.-W.,  {Ren} J.,  {Yang} J.,  {Zou} Z.-C.,   {Zhu} J.-P.,
  2024, \mn@doi [\apjl] {10.3847/2041-8213/ad2df6}, \href
  {https://ui.adsabs.harvard.edu/abs/2024ApJ...964L...9W} {964, L9}

\bibitem[\protect\citeauthoryear{{Xiao} et~al.,}{{Xiao}
  et~al.}{2022}]{2022arXiv220502186X}
{Xiao} S.,  et~al., 2022, \mn@doi [arXiv e-prints] {10.48550/arXiv.2205.02186},
  \href {https://ui.adsabs.harvard.edu/abs/2022arXiv220502186X} {p.
  arXiv:2205.02186}

\bibitem[\protect\citeauthoryear{{Yang} et~al.,}{{Yang}
  et~al.}{2015}]{2015NatCo...6.7323Y}
{Yang} B.,  et~al., 2015, \mn@doi [Nature Communications] {10.1038/ncomms8323},
  \href {https://ui.adsabs.harvard.edu/abs/2015NatCo...6.7323Y} {6, 7323}

\bibitem[\protect\citeauthoryear{{Zhang}, {Yi}, {Zhang}, {Xiong}  \&
  {Xiao}}{{Zhang} et~al.}{2022}]{2022ApJ...939L..25Z}
{Zhang} Z.,  {Yi} S.-X.,  {Zhang} S.-N.,  {Xiong} S.-L.,   {Xiao} S.,  2022,
  \mn@doi [\apjl] {10.3847/2041-8213/ac9b55}, \href
  {https://ui.adsabs.harvard.edu/abs/2022ApJ...939L..25Z} {939, L25}

\bibitem[\protect\citeauthoryear{{Zou} \& {Cheng}}{{Zou} \&
  {Cheng}}{2024}]{2024ApJ...973..126Z}
{Zou} L.,  {Cheng} J.-G.,  2024, \mn@doi [\apj] {10.3847/1538-4357/ad6dd9},
  \href {https://ui.adsabs.harvard.edu/abs/2024ApJ...973..126Z} {973, 126}

\bibitem[\protect\citeauthoryear{{van der Klis}}{{van der
  Klis}}{1989}]{1989ARA&A..27..517V}
{van der Klis} M.,  1989, \mn@doi [\araa]
  {10.1146/annurev.aa.27.090189.002505}, \href
  {https://ui.adsabs.harvard.edu/abs/1989ARA&A..27..517V} {27, 517}

\makeatother
\end{thebibliography}

%%%%%%%%%%%%%%%%% APPENDICES %%%%%%%%%%%%%%%%%%%%%
\appendix

%%%%%%%%%%%%%%%%%%%%%%%%%%%%%%%%%%%%%%%%%%%%%%%%%%
\bsp	% typesetting comment
\label{lastpage}
\end{document}